\documentclass[a4paper,11pt]{article}
\pdfoutput=1

\usepackage{jheppub}
\usepackage{amsmath,amssymb,braket,bm,hyperref,tikz,tikz-feynman,mathbbol,xcolor,mathtools,bbm,soul}

\usepackage{graphicx} % Required for inserting images
\usepackage{epsfig}
\usepackage{breakurl}
\usepackage{color}
\usepackage{tikz}
\usepackage[utf8]{inputenc}
\usetikzlibrary{snakes}
\usetikzlibrary{decorations}
\usetikzlibrary{trees}
\usetikzlibrary{decorations.pathmorphing}
\usetikzlibrary{decorations.markings}
\usetikzlibrary{external}
\usetikzlibrary{intersections}
\usetikzlibrary{shapes,arrows}
\usetikzlibrary{arrows.meta}
\usetikzlibrary{calc}
\usetikzlibrary{shapes.misc}
\usetikzlibrary{decorations.text}
\usetikzlibrary{backgrounds}
\usetikzlibrary{fadings}
\usetikzlibrary{tikzmark,calc,arrows,shapes,decorations.pathreplacing}

\def\dd{\mathrm{d}}
\def\hd{\hat{\mathrm{d}}}
\def\hdelta{\hat \delta}

\def\helicity{{\eta}}
\def\Smatrix{\mathbbm{S}}
\def\SoftTO#1#2#3{{\mathcal{A}(#1, #2 | #3)}}
\def\SoftOT#1#2#3{{\mathcal{A}(#1 | #2, #3)}}
\def\sfq{\mathsf{q}}
\def\sfQ{\mathsf{Q}}
\def\SoftCharge#1{{Q_s[#1]}}

\def\DressedStateR#1#2{{|{#1}[{#2}]\rangle}}
\def\DressedStateL#1#2{\prescript{}{}{{}}\langle {#1}[{#2}]|}

\def\DressedA#1{{A[#1]}}
\def\DressedPhi#1{{\phi[#1]}}

\newcommand{\nhat}{{\bm n}}
\newcommand{\Khat}{\hat{\bm K}}
\newcommand{\Kvec}{{\bm K}}

\newcommand{\Sphere}{S^2}
\newcommand{\nK}{{\rm K}}

\renewcommand{\[}{\begin{equation}\begin{aligned}}
\renewcommand{\]}{\end{aligned}\end{equation}}

\renewcommand{\v}[1]{\bm{#1}}
\definecolor{allOrderBlue}{rgb}{0.4,0.5,1}
\usepackage{subcaption}
\DeclareMathOperator{\timeOrder}{\mathcal{T}}

\title{
Supertranslations are Soft Dressings
}

\author[a, b]{Asaad Elkhidir}
\author[a]{Donal O'Connell}
\author[c,d,e]{Radu Roiban}

\affiliation[a]{Higgs Centre for Theoretical Physics, School of Physics and Astronomy, The University of Edinburgh, Edinburgh EH9 3JZ, Scotland, UK}
\affiliation[b]{Institut des Hautes Études Scientifiques, 91440 Bures-sur-Yvette, France}
\affiliation[c]{Institute for Theoretical Studies, ETH Zurich, 8092 Zurich, Switzerland}
\affiliation[d]{Institute for Gravitation and the Cosmos,
Pennsylvania State University,
University Park, PA 16802, USA}
\affiliation[e]{
Institute for Computational and Data Sciences,
Pennsylvania State University,
University Park, PA 16802, USA}

\emailAdd{elkhidir@ihes.fr}
\emailAdd{donal@ed.ac.uk}
\emailAdd{radu@phys.psu.edu}

\abstract{
Asymptotic states of massive particles in electrodynamics and gravity are not uniquely defined because of the absence of a mass gap.
Working in the KMOC formalism, we study the classical implications of this freedom in the dressing of massive particle states.
We show that coherent-state dressings modify observables by a large gauge/BMS transformation.
The associated coherent-state displacement operator can be completed into a conserved charge, the BMS charge in the gravitational case, which implements large gauge transformations on scattering data.
The parameter of our dressing is directly the large gauge/BMS transformation parameter, and we explain that observables do not depend on monopolar and dipolar parts of the transformation parameter to all orders.
Our construction is driven by soft factorization rather than by an underlying local symmetry, and we illustrate this by exhibiting the same structure in a scalar theory with no gauge symmetry. 
We explore the action of large gauge transformations on two-particle states for scalar, abelian gauge and graviton mediators, obtaining a universal shift of the 
impact parameter and clarifying the frame dependence of the radiated angular momentum vis \`a vis the frame-independence of the total angular momentum.
The transformation connecting the canonical BMS frame, in which the Schwarzschild metric takes the Kerr-Schild form, to the intrinsic frame, in which it takes the De Donder form, illustrates these effects concretely.
}

\begin{document}

\addtocontents{toc}{\protect\setcounter{tocdepth}{2}}

\maketitle

\section{Introduction}

It is now a decade since the first detection of gravitational waves. 
Since the initial breakthrough, we have observed over 300 events~\cite{LIGOScientific:2026sit, LIGOScientific:2026wfs}, including the spectacular highlight of the binary neutron star merger GW170817 with its associated kilonova.
Upcoming improvements in gravitational-wave detectors~\cite{Punturo:2010zz, LISA:2017pwj, Reitze:2019iox, LIGOasharp, Abac:2025saz} promise dramatic gains in both sensitivity and frequency coverage in the next decade. Realizing the full scientific potential of these observations, however, demands high-precision theoretical modeling across a large parameter space of the main observable: the waveform.

In addition to classic approaches such as numerical relativity~\cite{Pretorius:2005gq, Campanelli:2005dd, Baker:2005vv, Damour:2014afa} and  the gravitational self-force (SF) program~\cite{Mino:1996nk, Quinn:1996am, Poisson:2011nh, Barack:2018yvs}, perturbative methods in quantum field theory, originally developed for particle collider applications, have been at the forefront of this quest for precision.
These include approaches based on effective field theory matching~\cite{Cheung:2018wkq}, eikonal-based resummations~\cite{DiVecchia:2020ymx, DiVecchia:2021bdo}, exponential representations of the $S$-matrix~\cite{Damgaard:2021ipf, Damgaard:2023ttc}, observable-centered frameworks~\cite{Kosower:2018adc}, heavy-particle effective theories~\cite{Damgaard:2019lfh}, EFTs tailored to extreme mass-ratio systems~\cite{Cheung:2023lnj, Kosmopoulos:2023bwc}, and classical worldline constructions~\cite{Kalin:2020fhe, Mogull:2020sak, Kalin:2022hph, Jakobsen:2022psy}. 
These advances have enabled a series of high-order post-Minkowskian (PM) calculations of inclusive two-body observables, particularly for non-spinning binaries~\cite{Bern:2019crd, Bern:2019nnu, Bern:2024adl, Bern:2021dqo, Bern:2021yeh, Driesse:2024xad, Driesse:2024feo, Jakobsen:2023hig, Jakobsen:2023ndj, Dlapa:2021vgp, Dlapa:2021npj, Bjerrum-Bohr:2021wwt, Bern:2025wyd, Driesse:2026qiz} and scattering waveforms~\cite{Herderschee:2023fxh,Elkhidir:2023dco, Georgoudis:2023eke, Georgoudis:2023lgf, Brandhuber:2023hhy, Bini:2024rsy}. 
(Effects due to spin, finite-size structure, tidal interactions, and the surrounding environment can also be incorporated in this framework; see, for example, Refs.~\cite{Blanchet:2013haa, Porto:2016pyg, Buonanno:2022pgc, Barausse:2014tra}.) 

Local (position-dependent) observables in gravitational theories are much less developed than in gauge theories, see e.g. \cite{Torre:1993fq,  Giddings:2005id, DeWitt:1967yk, Donnelly:2015hta, Bergmann:1961wa,Cheung:2026euf} for classic and more recent discussions of this interesting and timely topic. 
In gauge theories, operators invariant under small gauge transformations are constructed by dressing charged fields with Wilson lines or Coulombic clouds~\cite{Dirac:1955uv, Mandelstam:1962mi, Lavelle:1995ty}, with gravitational counterparts developed in Refs.~\cite{Torre:1993fq,  Giddings:2005id, DeWitt:1967yk, Donnelly:2015hta, Bergmann:1961wa,Cheung:2026euf}.
The gravitational waveform, i.e. the radiative degrees of freedom of the spacetime metric near future null infinity, is the primary observable of gravitational-wave astronomy. 
From a theoretical perspective, however, it is known that the current definition yields a scattering waveform which exhibits infrared divergences~\cite{Herderschee:2023fxh,Elkhidir:2023dco,  Georgoudis:2023lgf, Brandhuber:2023hhy}, and transforms non-trivially under large gauge transformations~\cite{Veneziano:2022zwh,Georgoudis:2023eke,Bini:2024rsy}, a freedom that operator dressing of the Wilson-line type does not fix. 
These two features are in fact related, and the gauge freedom assists with the removal of infrared divergences and, as we will see, with the apparent consequences of arbitrarily-soft initial-state radiation.
Constructing local observables that are infrared safe and invariant under both large and small gauge transformations remains an interesting open problem.

We will be particularly interested in the large gauge transformations known as supertranslations; these act on the retarded time $u$ as 
\[
u \rightarrow u + T(\theta, \phi) \,
\label{eq:u_shift}
\]
where $T(\theta, \phi)$ is an arbitrary function of angles. Simultaneously, the shear --- i.e. the $1/r$-suppressed correction to the metric on the celestial sphere --- is modified by a certain two-derivative operator acting on the supertranslation parameter $T(\theta, \phi) $. The latter does not affect the radiative Newman-Penrose scalar $\psi_4(u)$, which instead transforms as~\footnote{The other Newman-Penrose scalars have more involved transformation rules, see e.g. Eq.~(6.86) of Ref.~\cite{Barnich:2016lyg}.}
\[
\psi_4(u, \theta, \phi) \rightarrow \psi_4(u + T(\theta, \phi),  \theta, \phi) \,.
\]
This fact raises several questions.
First, is this BMS transformation of the gravitational waveform observable?
If it is, what is the physically relevant choice of $T(\theta, \phi)$ for the data?
How can we incorporate this choice of ``frame'' (that is, a specific choice of supertranslation parameter $T$) in our computation of the waveform using quantum field theory methods?

We will address the last of these questions, leaving the others for the future. Supertranslations lead to an ambiguity in the classical waveform. We will explain how this is related to another ambiguity of scattering theory: in theories without a mass gap, the single-particle pole of a charged (massive) particle's propagator is not well-defined. 
There is a complementary extensive literature that uses general relativity methods to explore the consequences of BMS transformations on the scattering matrix and their connection to infrared singularities, see e.g.~\cite{Strominger:2013jfa, Strominger:2014pwa, Campiglia:2014yka, He:2014laa, Pasterski:2015tva, Campiglia:2015yka, Campiglia:2015qka, Conde:2016csj, Kapec:2015vwa, Henneaux:2018cst, Kapec:2017tkm, Himwich:2020rro, Campiglia:2015kxa} and the reviews~\cite{Strominger:2017zoo, Ashtekar:2018lor, Raclariu:2021zjz, Pasterski:2021rjz}.

Long-range interactions, mediated by massless fields, spoil the assumptions of ordinary scattering theory: the early- and late-time motion of charged particles is not free, and the interaction leaves a persistent imprint even asymptotically. A familiar symptom is the infrared divergence of the scattering matrix of ideal plane-wave states. 
A standard option is to restrict attention to IR-safe observables. 
An alternative, suggested by the structure of the two-point function of charged (massive) fields, is to build the asymptotic dynamics into the definition of the asymptotic states by dressing them with coherent states of low-energy mediators, i.e. massless scalars, photons or gravitons.
Indeed, the K\"all\'en--Lehmann representation of these two-point functions shows that, without a mass gap, the single-particle pole is not isolated: it sits at the edge of the continuum of states containing arbitrarily soft mediators. There is therefore no preferred split between a charged particle and the same particle dressed with additional soft quanta. 
The extreme case of this freedom is dressing by strictly \emph{zero}-energy mediators. We will see below that this corresponds to the idealized case of the dressing being defined at strict null infinity. The Coulomb-like field of a static particle is time-independent and is therefore, in momentum space, a condensate of zero-frequency quanta; states differing by such a dressing describe the same particle with different accompanying static fields. 
It is precisely this zero-energy freedom that we will identify with the choice of BMS frame.

The structure of the resulting space of states has been studied extensively~\cite{Mirbabayi:2016axw, Bousso:2017dny, Flanagan:2022pmj, Satishchandran:2019pyc, Prabhu:2022zcr, Strominger:2014pwa, Strominger:2017zoo, He:2014laa}. While the Hilbert space can be factorized as
\[
\Gamma = \Gamma_s \otimes \Gamma_h,
\label{factorization0}
\]
with $\Gamma_s$ containing the zero-energy states and $\Gamma_h$ their finite-energy complement, no choice of finite-energy states makes time evolution preserve this factorization, at least in QED~\cite{Flanagan:2022pmj}. The minimal allowed evolution mixes the two factors,
\[
[s, h]\mapsto [{\bar s}(s), {\bar h}(s, h)] \, ,
\label{factorization1}
\]
so that the final state of the finite-energy particles depends on the zero-energy content of the initial state. Our results provide a concrete realization of a more general evolution,
\[
[s, h]\mapsto [{\bar s}(s,h), {\bar h}(s, h)] \, .
\label{factorization}
\]
The zero-energy mediator dressing of the initial state, i.e. the choice of asymptotic frame, enters observables of the hard particles, while the memory of the waves emitted by the hard particles alters the zero-energy mediator content of the final state. We will also see that the soft contributions to the hard sector evolution can be absorbed by a reparametrization of local observables. 
We do not attempt here a general analysis of this factorization problem. 

Asymptotic on-shell mediators, here a generic field $\Psi_M(x)$, are created by the perturbative operator~\cite{Cristofoli:2021vyo}
\begin{equation}
\begin{aligned}
\label{eq:statPhase}
\Psi_{M}(x) &= \sum_\helicity \int \dd\Phi(k) \, e_\helicity{}_{M}(k) a^\helicity(k) \, e^{-i k \cdot x} + \textrm{h.c.} 
\\
&\simeq \sum_\helicity\frac{-i}{4\pi |\v{x}|} \int_0^\infty  \dd \omega\, e_\helicity{}_{M}(k) a^\helicity(k) e^{-i \omega u} |_{k = \omega n} + \textrm{h.c.} +{\cal O}(1/|\bm x|^2)\,,
\end{aligned}
\end{equation}
where $M$ is a multi-index appropriate for the Lorentz representation of $\Psi(x)$, $u = x^0 - |\v x|$ is the retarded time and 
\[\label{eq:defOfn}
n^\mu = (1, \hat{\v{x}}) \,.
\]
The integration on the first line of equation~\eqref{eq:statPhase} is over the on-shell phase space of the massless mediator, so that
\begin{align}
\dd\Phi(k) \equiv \hd^4 k \, \hdelta_+(k^2)\,, \quad  \hd k \equiv \frac{\dd k}{2\pi} \,, \quad \hdelta(k) \equiv 2 \pi \, \delta(k) 
\, , \quad
\delta_+(k^2) \equiv \delta(k^2) \Theta(k^0) \, .
\end{align}
The second line of Eq.~\eqref{eq:statPhase} follows using a standard stationary phase approximation which follows from the assumption that $\omega |\v{x}| \gg 1$ in the exponent on the first line, see e.g. Ref.~\cite{Cristofoli:2021vyo}.
Sufficiently-soft mediators, however, lead to a breakdown of this assumption.
Even in the strict asymptotic limit, $|\v{x}|\rightarrow\infty$, excitations with $\omega=0$ can lead to a departure from $\omega |\v{x}| \gg 1$.

Such an asymptotic limit, and consequently the corresponding $\omega=0$ excitations, are of course idealizations. At any finite and large distance an analogous breakdown occurs, for mediators with frequencies smaller than the distance to the observer, $\omega < 1/|\v{x}|$. 
In fact, states with zero and nearly-zero-energy mediators are ubiquitous. 
Take, e.g. the gravitational waveform sourced when two point-like objects scatter. The difference between the metric in the far future and the far past is non-zero, because the different initial and final velocities of the objects lead to different linearised Schwarzschild solutions describing the individual particles. Correspondingly, the metric perturbation at fixed large distance $|\v x|$ will contain modes of frequency of order $1/|\v{x}|$ or smaller. 
In the same spirit, consider a single particle produced through some gravitational process. The particle's parents must have experienced some acceleration, which in turn produced some gravitational radiation. The memory of these waves, which is a static field and thus a condensate of physical zero-energy gravitons, will accompany the resulting particle.

It is therefore of interest to develop a formalism that accounts for these static or effectively static fields and their interactions with finite-energy particles. 
We will do this by dressing the initial states by suitable coherent states of mediators.
The shape function of the dressing is largely free, and different choices have traditionally been distinguished by their virtues as IR regulators — beginning with the construction of Kulish and Faddeev, see e.g.~Refs.~\cite{Carney:2018ygh, Lippstreu:2025jit, Chicherin:2025keq} for recent discussions. 
We take a different point of view: a coherent-state dressing whose shape is supported on zero-energy mediators changes the time-independent expectation value of the mediator field at null infinity. This expectation value defines the on-shell Coulomb field of the particle\footnote{We may interpret this as the on-shell version of the suggestion in Ref.~\cite{Caron-Huot:2023vxl} of incorporating the effects of low-frequency modes by including the \emph{off-shell} Coulomb field of the initial particles.},
which we also identify with the asymptotic (symmetry) frame.
We connect this dressing freedom on the one hand with the lack of a unique definition of charged states in the presence of massless mediators  and on the other with the choice of BMS frame in gravity, or its analogue in theories with other massless mediators. 
We will show in full generality that BMS supertranslations correspond to dressings of asymptotic states with zero-energy gravitons.

This identification has two complementary interpretations. On the one hand, the on-shell three-point amplitudes with real kinematics carry distributional support on configurations with exactly zero-energy mediators, and the choice of dressing corresponds to a particular organization of this distributional content. This is the perspective taken in Ref.~\cite{Elkhidir:2024izo}.
On the other, one may declare that the $S$-matrix has no on-shell three-point amplitudes with real momenta and choose a coherent state dressing to fix the desired asymptotic frame. In this paper we will take this perspective.
Since, as we show, BMS supertranslations (and more generally large gauge transformations) are equivalent to changing the coherent state dressing, it follows that the distributional content of three-point amplitudes with real kinematics can be changed by such transformations\footnote{It would be interesting to explore the consequences of these distributional three-point amplitudes on higher-point amplitudes via the BCFW on-shell recursion relations. We note, moreover, that both these as well as three-point amplitudes with an off-shell graviton can be read off the asymptotic metric linearized around Minkowski spacetime.}.

Two asymptotic frames are distinguished in gravitational-wave physics: the intrinsic frame of the Multipolar-Post-Minkowskian (MPM) formalism~\cite{Blanchet:1985sp, Blanchet:1989ki, Blanchet:2013haa} and the canonical frame natural to the QFT/EFT approach. We show, echoing Ref.~\cite{Bonga:2018gzr}, that these have a sharp geometric meaning: the intrinsic frame is the asymptotics of the linearised Schwarzschild metric in de~Donder coordinates, while the canonical frame is the asymptotics of the same metric in Kerr-Schild coordinates. 
The supertranslation connecting them, identified in Ref.~\cite{Veneziano:2022zwh}, is a specific zero-energy graviton dressing in our construction, see also~\cite{Menezes:2026edi,Menezes:2026fmj}. This suggests that classical perturbation theory organized directly around Kerr-Schild data, rather than around de~Donder gauge with a subsequent supertranslation, may be a more natural starting point for scattering observables. We return to this point in the conclusions.

The most physically transparent consequence of this construction is a shift of the impact parameter under changes of the particles' Coulomb field or, equivalently, under changes of the asymptotic frame. 
A central object in our construction is a conserved charge $Q=Q_s+Q_h$, in which the soft piece $Q_s$ is the dressing phase connecting the two frames and the hard piece $Q_h$ is fixed by the requirement that the total charge $Q$ is invariant under time evolution, $[S,Q]=0$.~\footnote{Our dressed states are not eigenstates of this $Q$ operator. This is in contrast with other coherent-state dressings~\cite{Choi:2019fuq, Choi:2017ylo, Choi:2018oel} which aim to construct asymptotic states which are eigenfunctions of the generator of large gauge transformations.  
} 
Through eikonal manipulations of the dressed final state, we then obtain an all-orders relation between the asymptotic impact parameter, the eikonal impact parameter, and the kinematics-dependent eigenvalue of $Q_h$ on the two-particle state.
This sharpens, by giving an explicit all-order formula, the observation of Ref.~\cite{Compere:2023qoa} (see in particular Sec. 9.4 there) that the impact parameter measured at infinity depends on the BMS frame.
As a consistency check, we verify at leading order that the change in mechanical angular momentum implied by our impact-parameter shift equals the negative of the radiated angular momentum's frame dependence — so that the total angular momentum is frame-independent. 
Thus, our all-orders expression for the change of the asymptotic impact parameter determines the radiated angular momentum's frame dependence to all orders in perturbation theory.
Specializing to the canonical $\leftrightarrow$ intrinsic supertranslation, we show that the corresponding leading-order impact-parameter shift reproduces the known frame dependence of the radiated angular momentum in general relativity~\cite{Bini:2022wrq} and in QED~\cite{Saketh:2021sri}. 
At the next order, our all-orders expression yields a prediction for the frame dependence of the angular-momentum loss that, to our knowledge, has not previously appeared in the literature.

The rest of the paper is organised as follows. 
In Section~\ref{sec:basicSetup} we review the structure of charged asymptotic states in theories with massless mediators and discuss dressed states and asymptotic frames, illustrating for the case of electromagnetism.
In Section~\ref{sec:dressedGravity} we discuss the analogous dressing in gravity, where it captures general BMS supertranslations, and identify the Kerr-Schild metric as the spacetime corresponding to the canonical BMS frame.
Section~\ref{sec:general_hard_charges} constructs the exact hard charge for an arbitrary supertranslation parameter, with the master integrals deferred to Appendix~\ref{app:int_general_dressing}. 
Section~\ref{sec:consequences} derives the all-orders relation between the asymptotic and eikonal impact parameters in the presence of a general dressing. 
Section~\ref{sec:VV} specialises to the canonical $\leftrightarrow$ intrinsic dressing for scalar, vector, and graviton mediators, computes the corresponding hard charges and impact-parameter shifts, and compares with existing results. 
Finally, section~\ref{sec:conclusions} summarises and discusses possible extensions of our work. 
Appendix~\ref{app:BMS} collects the rudiments of BMS that we use, and 
Appendix~\ref{app:int_general_dressing} contains the evaluation of the master integrals.

%%%%%%%%%%%%%%%%%%%%%%%%%%%%%%%%%%%%%%%%%%%%%%%%%%%%%%%%%%%%%%%
%%%%%%%%%%%%%%%%%%%%%%%%%%%%%%%%%%%%%%%%%%%%%%%%%%%%%%%%%%%%%%%

\section{General setup}
\label{sec:basicSetup}

An important ingredient in the calculation of observables in quantum field theory is the definition of asymptotic states. 
However, in theories with massless interacting particles, these states cannot be defined uniquely because of the existence of arbitrarily low-energy modes of mediator fields. 
In this section we review these aspects of asymptotic states
and argue that the choice of (infrared content of the) single-particle states defines the asymptotic (symmetry) frame.\footnote{This asymptotic frame fixes the asymptotic symmetries when the latter are present. The considerations below apply equally well when there are no asymptotic symmetries, e.g. in gravitational theories when Bondi coordinates cannot be reached or in theories with massless scalar fields without a shift symmetry. For this reason we refer to ``asymptotic frames'' rather than to ``asymptotic symmetry frames'' below.}
In order to discuss the physical concepts in a mathematically simple setting we largely focus on the electromagnetic case in this section.
The gravitational case will be the topic of Section~\ref{sec:dressedGravity}.

\subsection{On asymptotic states of charged particles}

Throughout this paper, we are interested in scattering processes involving interacting massive particles. 
It is often convenient to introduce ladder (creation and annihilation) operators for a massive particle, which we will denote as $A^\dagger(p)$ and $A(p)$.
These satisfy the familiar commutation relations
\[\label{eq:basicLadderCRs}
&[A(p), A(p')] = 0 \,, \\
&[A(p), A^\dagger(p')] = \delta_\Phi(p, p') \,, \\
&[\mathbb{P}^\mu_\text{tot}, A(p)] = -p^\mu \, A(p) \,,
\]
where $\delta_\Phi(p,p')$ is the on-shell delta function
\[
\delta_\Phi(p, p') = 2 E_p \, \hat \delta^3(\v{p} - \v{p}') \,,
\]
and $\mathbb{P}^\mu_\text{tot}$ is the total momentum operator. 
The single-particle states are then
\[
\ket{p} = A^\dagger(p) \ket{0} \,.
\]
The quantum field $\phi(x)$ associated with these particles can be written in the usual way in terms of these ladder operators.

The time-ordered two-point function of this (renormalized) field $\phi(x)$, in an interacting relativistic quantum field theory satisfying the usual axioms\footnote{That is, Poincar\'e invariance, 
positivity/unitarity, 
existence of a vacuum, and 
existence of a physical ($P^0>0, P^0\ge |\bm P|$) spectrum of the energy-momentum operator $P^{\mu}$ in the closed forward light cone.
},
admits the spectral representation
\begin{equation}
    i\Delta_F(p)\equiv \int \dd^4x\,e^{ip\cdot x}\,\langle 0|\timeOrder \phi(x)\phi(0)|0\rangle
    \;=\;\int_{0}^{\infty} \dd s\,
    \frac{\rho(s)}{p^2-s+i\epsilon} \ .
    \label{eq:KL}
\end{equation}
The symbol $\timeOrder$ above is the usual time-ordering operator;
the spectral density $\rho(s)$ is non-negative for gauge-invariant local fields\footnote{
For gauge-variant fields in covariant gauges, positivity can fail. Our discussion here focuses on charged matter, whose long-range fields have complicated asymptotics in the presence of massless mediators.}.
The formal definition of $\rho(s)$ is 
\begin{align}\label{eq:undressedSpectralDensity}
\rho(s) = \sum_n |\langle p, n|\phi(0)|0\rangle|^2 \delta(s-m_n^2) \ ,
\end{align}
where the index $n$ runs over all the single- and multi-particle states of total momentum $p$ of the theory, and the matrix element includes an implicit integration over the relative momenta of their constituents.
This representation was established by K{\"a}ll{\'e}n and Lehmann~\cite{Kallen:1952zz, Lehmann:1954xi} and is reviewed in many modern texts. The key structural point is that the exact propagator
is a superposition of free propagators with all possible invariant masses $s^{1/2}$, weighted by $\rho(s)$.

In a theory with a mass gap and only massive stable particles, the long-distance dynamics is effectively ``short-ranged''; one can typically construct scattering states leading to an $S$-matrix that is well-defined on a Fock space of massive particles.
In such a theory, the field $\phi$ has overlap with a stable one-particle state of mass $m$ (and creates only such states at asymptotic times and distances) and the spectral density contains a delta function,
\begin{equation}
    \rho(s)= Z\,\delta(s-m_\text{phys.}^2) + \rho_{\rm cont}(s) \Theta(s- m_M^2) \ ,
    \label{eq:massiverho}
\end{equation}
which is located at the physical mass $m_\text{phys.}$ of the one-particle state. Its coefficient $Z$ is the wave-function renormalization constant of $\phi$, and $\Theta$ is the Heaviside step function.
In a purely massive theory, the threshold $s = m_M^2$ marking the beginning of the continuum part of the spectral density corresponds to the lowest multi-particle invariant mass squared. The mass gap $m_M^2>m_\text{phys.}^2$ ensures that the single-particle pole is isolated. 
This separation between the single-particle pole and the continuum allows for a clear definition of the physical mass and of the asymptotic states, and thus it is key to the LSZ reduction and its refinements~\cite{Collins:2019ozc}. 

The situation changes qualitatively when the massive particles are charged under massless mediator fields or, more generally, in the presence of massless particles. 
The mediators then produce long-range fields which are the origin of two intertwined issues.

On the one hand, the usual $S$-matrix, with standard Fock states as asymptotic states, exhibits infrared divergences, see e.g.~\cite{Yennie:1961ad, Weinberg:1965nx}, whose origin is the exchange of arbitrarily-low-energy quanta. The structure of these divergences are well-understood in gauge and gravitational theories, as is their fate in sufficiently-inclusive observables~\cite{Bloch:1937pw, Lee:1964is, Kinoshita:1962ur}. 
On the other, there are no naive Fock asymptotic states for charged particles, see e.g.~\cite{Schroer:1963gw, Frohlich:1979xp, Buchholz:1986uj}. Indeed, charged particles are always accompanied by mediator quanta of arbitrarily low energy. Physically, acceleration/emission processes populate the soft sector, while mathematically Gauss's law ties the charge to an ${\cal O}(1/r^2)$ flux at infinity (in four dimensions).\footnote{The situation is more involved in cases where the mediator field itself is measurable (not only its field strength), as in gravitational theories. } 
This obstructs a sharp one-particle description for charged states.
In algebraic terms, Buchholz~\cite{Buchholz:1986uj} showed that states carrying a charge measurable via the Gauss law cannot be eigenstates of the mass operator; the charged sector contains states dressed with some configuration of essentially-zero energy excitations rather than standard one-particle states.

These features have a reflection in the K{\"a}ll{\'e}n-Lehmann representation of the two-point function of charged fields. 
Intuitively, a charged field creates a superposition of states with one charged particle and an arbitrary configuration of soft mediators. 
Since soft mediators can carry arbitrarily small energy, there is no gap between the charged ``one-particle" configuration and the continuum, and thus the spectral density $\rho(s)$ cannot exhibit a separated delta-function. 
Effectively, the residue of the one-particle pole vanishes: $Z=0$.
The absence of such a pole is sometimes described as the breakdown of the standard LSZ picture for charged excitations, see e.g.~\cite{Kulish:1970ut, Dybalski:2017mip, Gass:2021idx}.
More formally, the terms in Eq.~\eqref{eq:massiverho} are labelled by the eigenvalues of the mass operator while, according to Ref.~\cite{Buchholz:1986uj}, charged states are not eigenstates of the mass operator.

In the rest of this section, we will link this feature of quantum field theories to a seemingly distinct ambiguity: large gauge transformations in classical field theories.

\subsection{Dressed asymptotic states} 
\label{sec:dressedState}

As we discussed, in theories with massless mediators genuine single-particle states do not exist.
Nevertheless, scattering theory with naive Fock single-particle states is extremely successful and the IR-divergences of $S$-matrix elements have been thoroughly understood~\cite{Yennie:1961ad, Weinberg:1965nx}.
Furthermore, dressed asymptotic states have been extensively used as a means to define IR-finite $S$-matrix elements, building on the foundational work of Kulish and Faddeev \cite{Kulish:1970ut}.
The Faddeev-Kulish proposal was to modify the asymptotic dynamics so that the in and out states are not bare charged particles, but rather are charged particles accompanied by a coherent state of soft mediators (photons, for example). 
Modern treatments and connections to soft theorems and asymptotic symmetries are reviewed in e.g. Ref.~\cite{Kapec:2017tkm}.

While we are not interested here in the construction of IR-finite $S$-matrix elements, it is nevertheless interesting to explore the interpretation of the Fock single-particle states dressed with various configurations of low-energy mediators and to understand their physical interpretation. 
One observation is that this dressing depends on the production mechanism of the specific state.
There are many ways of producing a given state, so the dressing --- which depends on the details of the production history --- is essentially arbitrary. 
A low-energy dressing of this type can be included even in the presence of an additional ``hard'' dressing, such as that defining (nonlocal) gauge-invariant asymptotic states for charged particles~\cite{Dirac:1955uv, Lavelle:1995ty,Bagan:1999jf,Cheung:2026euf}.
Physical observables (e.g. inclusive cross sections, gauge-invariant correlators) are gauge independent, but the \emph{partition} between ``hard charged particle'' and ``soft cloud'' can be gauge- and scheme-dependent.

A large class of dressed states are those dressed with coherent states of massless mediators: 
\begin{equation}
%    |{p}\rangle^{\rm dressed}
%\DressedStateR{\psi}
%     \;\sim\;
    \exp\!\left[i\int\! \dd\Phi(k, p)\, A^\dagger(p) A(p) \,
      \bigl(F^\helicity_p(k)\,a_\helicity^\dagger(k)+\bar{F}_p{}_\helicity(k)\,a^\helicity(k)\bigr) \right]\,|p\rangle \ ,
\label{eq:coherent_dressing}
\end{equation}
where $\helicity$ labels the mediator helicity.
The ``shape'' functions $F^\eta_p(k)$ are arbitrary, and we have denoted the conjugate shape function by $\bar{F}_p{}_\helicity(k) = (F_p^\helicity(k))^*$. 
The shape functions can be chosen on physical grounds to satisfy various conditions. Examples are that  
the states satisfy some physical requirements such as the Gauss law; 
or that the states possess some desired radiation content; 
or that the resulting $S$-matrix elements enjoy desirable properties;
or that the time-independent long-range field of a particle, which we refer to as the Coulomb field, has a desired form. 
This last point hinges on the coherent states having a classical limit.

From the perspective of the K{\"a}ll{\'e}n-Lehmann representation of the two-point function of charged fields, a successful dressing reorganizes which states are treated as ``one-particle'' asymptotic configurations.
Rather than expecting an isolated pole in the two-point function of a \emph{bare} charged field, one constructs dressed asymptotic states whose long-distance behaviour better matches the true charged sector. 
However, because the long-range field is physical and extends to infinity, the dressed excitations do not behave like strictly local particles, as can be clearly seen by interpreting Eq.~\eqref{eq:coherent_dressing} in position space.

In the strict asymptotic limit, in which the particle is free, the Coulomb field of the particle is time-independent, so its momentum-space Fourier transform is localized at zero frequency. 
Thus, we expect that the Coulomb field can be captured by a shape function $F^\helicity_p(k)$ with support on zero-energy mediators, 
\[
F^\helicity_p(k)= \hdelta(\omega) f^\helicity_p(n) \,,
\label{localization}
\]
where $k^0 = \omega$ and $k^\mu = \omega n^\mu$.
Of course, the strict asymptotic limit, and consequently the localization \eqref{localization}, are only theoretical constructs. In real scattering experiments the asymptotic states are prepared and observed at finite distances and times. Consequently, $F^\helicity_p(k)$ really spans a range of low energies below the resolution of any measuring apparatus: these are, in fact, infrared dressings.
Physically, the $\omega=0$ configuration should be defined as a limit of some sequence of finite energy configurations.
This perspective is very close to the spirit of~\cite{Feal:2022iyn,Feal:2022ufw} which emphasise the importance of the order of limits for infrared divergences and particle dressing in a worldline approach to scattering amplitudes.
Although the localization~\eqref{localization} is highly idealised, it is nevertheless theoretically convenient, and we will refer to these localized dressings as ``zero-energy'' dressings.
We note in passing that the discussion of Ref.~\cite{Elkhidir:2024izo} indicates that we may interpret our  on-shell, zero-energy dressing $f_p^\helicity$ 
as a change in the on-shell three-point amplitudes of the theory. 
It is also worth emphasising that, in view of the stationary phase approximation~\eqref{eq:statPhase}, the null vector $n^\mu$ which appeared in equation~\eqref{localization} is precisely the vector of equation~\eqref{eq:defOfn}.

It is necessary to ensure that interactions of Fock states with low energy particles are not over-counted. To this end we need to introduce a separation between the mediators included in the (coherent) state dressing and those included in the scattering matrix with external charged Fock states.
In Ref.~\cite{Elkhidir:2024izo}, which worked with states defined at strict asymptotic infinity, we showed that it is sufficient to exclude all three-point elements with exactly zero-energy mediators from the $S$ matrix.
It is perhaps less clear what the appropriate separation is in the presence of a finite-energy cutoff on the energy of mediators included in the state dressing, especially when the asymptotic states are defined away from strict infinite distance. 
In the following, as in Ref.~\cite{Elkhidir:2024izo}, we will define states as asymptotic infinity; while we interpret the vanishing-energy mediators as a limiting configuration, we will assume that the cutoff is much smaller than any other scale in the problem, and moreover that the limit is taken as soon as all commutators have been evaluated.

The operator~\eqref{eq:coherent_dressing} must be a Lorentz scalar to preserve covariance of the state, and therefore we must balance the helicity transformations of the ladder operators and of the shape functions.
Let us write the helicity transformation of the annihilation operator through an angle $\theta$ as
\[
a^\helicity(\omega, n) \rightarrow e^{i \helicity \theta} a^\helicity(\omega, n) \,.
\]
Lorentz covariance requires the localized shape function $f^\helicity_p(n)$ to transform in precisely the same way:
\[
f^\helicity_p(n)  \rightarrow e^{i \helicity \theta}  f^\helicity_p(n) \,.
\]
This can be achieved by setting
\[\label{eq:shapeAsDerivative}
f^\helicity_p(n) = \bar{\epsilon}^\eta{}^{\mu}(n) \frac{\partial}{\partial n^\mu} \, T_p(n) \,,
\]
for real $T_p(n)$; this choice will be sufficient for our purposes\footnote{We restrict here to the case of a spin 1 mediator; spin 0 is a straightforward exercise. We will discuss the spin 2 case in detail below in section~\ref{sec:dressedGravity}.}.
We will discuss the rationale for this choice in more detail in the next subsection. 

We therefore write our dressed states
in terms of a Hermitian operator $Q_s$, which we will refer to as the ``soft charge''.
The soft charge is defined explicitly by 
\[
\SoftCharge{T} =  \sum_\helicity\int\! \dd\Phi(k, p)\,
      A^\dagger(p) A(p) \hdelta (\omega) \bigl(\bar{\epsilon}^\helicity \cdot \partial \, T_p(n)\,a_\helicity^\dagger(k)+\epsilon_\helicity \cdot \partial \, T_p(n) \,a^\helicity(k)\bigr) \Big|_{k = \omega n}\,.
\label{eq:Qc}      
\]
The construction of dressed single-particle states extends naturally to single-particle wave packets. 
The dressing of a wavepacket $|{\psi}\rangle$ is simply 
\begin{align}
\label{eq:dressing}
|{\psi}\rangle \rightarrow 
%|{\psi}\rangle{}^\text{dressed} 
\DressedStateR{\psi}{T} \coloneqq \int \dd\Phi(p) \, \varphi(p) \, e^{i \SoftCharge{T}} |{p}\rangle 
= e^{i \SoftCharge{T}}
|{\psi}\rangle 
%\eqqcolon U[T] |{\psi}\rangle 
\, .
\end{align}
We will assume that the bare wavepacket $|{\psi}\rangle$ is normalized, so the dressed one is also normalized.

\subsection{Asymptotic frames}
\label{sec:frame}

We now develop an understanding of the classical implications of our dressings by studying the expectation value of the electromagnetic field in dressed wavepacket states.
In fact, we will see that one can choose $T_p(n)$ so that the expectation value of $\mathbbm{A}^{\mu}(x)$ reproduces the Coulomb field of the particle in any desired asymptotic frame.

First, we must discuss the quantum field itself. Since we are interested in the fields measured by a distant observer, we use the stationary phase approximation~\eqref{eq:statPhase}, specialised to massless vector particles, to write the quantum field as
\[\label{eq:photonStatPhase}
\mathbbm{A}^{\mu}(x) \simeq \sum_\helicity\frac{-i}{4\pi |\v{x}|} \int_0^\infty  \dd \omega\, \epsilon_\helicity{}^{\mu}(k) a^\helicity(k) e^{-i \omega u} \big|_{k = \omega n} + \textrm{h.c.} +{\cal O}(1/|\bm x|^2)\,.
\]
The stationary phase approximation is justified when $\omega |\v{x}|$ is large. 
Consistent with our earlier discussion of the idealisation that $|\v {x}| \to \infty$, we \emph{define} the photon field at large distance by Eq.~\eqref{eq:photonStatPhase} even for small/zero frequency photons.

First let us take a moment to clarify in what sense we expect the expectation value of the quantum field $\mathbbm{A}^{\mu}(x)$ to reproduce expressions familiar from classical field theory.
To that end, it is helpful to specify the coordinates $x^\mu$ in more detail, in particular by connecting the observation point to the vector $n^\mu$ appearing in equation~\eqref{eq:photonStatPhase}.
As in section~\ref{sec:dressedState} we denote the proper velocity of the observer as $U^\mu$, we may  write the observer's position as
\[
\label{eq:coordChoice}
x^\mu = u U^\mu + r n^\mu \,.
\]
In this expression, the scalar $u$ may be interpreted as the retarded time, while $r = | \bm x|$ is the radial distance in the observer's rest frame.
Without loss of generality we may choose $n \cdot U = 1$, and the vector $n^\mu$ takes precisely the form given in equation~\eqref{eq:defOfn}.
We can then view the coordinates as $u$, $r$ and the two angles $\theta^A$ for $A=1,2$ which parametrise $n^\mu$.
As a differential form we then have
\[\label{eq:dcoords}
\dd x^\mu &= U^\mu \, \dd u + n^\mu \, \dd r + r \, \dd n^\mu \\
&= U^\mu \, \dd u + n^\mu \, \dd r + r  \,\partial_A n^\mu \, \dd \theta^A \,.
\]

We assume that our observer is capable of measuring the expectation value $\braket{\mathbbm{A}^\mu(x)}$ of the field\footnote{While this is unphysical for QED, where one measures field strengths and reconstructs the vector potential as desired, we make this assumption to stay close to gravity, where one measures displacements or, equivalently, (the expectation value of) the metric fluctuation (operator).}.
It is then very natural to choose the gauge of the polarisation vectors so that $U \cdot \epsilon_\helicity = 0$. 
With these choices, the one-form version of the field takes an instructive form:
\[
\braket{\mathbbm{A}(x)} \cdot \dd x = \dd \theta^A \,
\sum_\helicity\frac{-i}{4\pi} \int_0^\infty  \dd \omega\,  \partial_A n^\mu \, \epsilon_\helicity{}_{\mu}(k) \, \braket{a^\helicity(k)} e^{-i \omega u} \big|_{k = \omega n} + \textrm{h.c.} +{\cal O}(1/r)\,.
\]
Note that the field expectation is purely angular: it involves $\dd \theta^A$ and not $\dd u$ or $\dd r$. 
It is then these angular components that we expect to match to classical field theory.
(Similar comments apply to the gravitational case.)

Now let us consider the particular expectation value on the dressed initial state~\eqref{eq:dressing}.
A straightforward calculation using equation~\eqref{eq:photonStatPhase} leads to
\[
\DressedStateL{\psi}{T} \mathbbm{A}(x)  \DressedStateR{\psi}{T} \cdot \dd x
= \frac{\dd n^\mu}{4\pi} \sum_\helicity \left( \epsilon_{\helicity}{}_{\mu} (n) \bar{\epsilon}^\helicity(n) \cdot \partial T_p(n) 
+ \bar{\epsilon}^\helicity_{\mu}(n) \epsilon_\helicity(n) \cdot \partial T_p(n)
\right) + \mathcal{O}(1/r)\,.
\label{eq:genericPhotonDressing}
\]
We perform the helicity sum in equation~\eqref{eq:genericPhotonDressing} using the completeness relation
\[
\sum_\eta \epsilon_{\helicity\mu} (n) \bar{\epsilon}^\helicity_{\nu} (n) = P_{\mu\nu} = - \big( \eta_{\mu\nu} - n_\mu U_\nu - n_\nu U_\mu + n_\mu n_\nu \big ) \,,
\label{eq:projectorDef}
\]
finding that
\[
\DressedStateL{\psi}{T} \mathbbm{A}(x)  \DressedStateR{\psi}{T} \cdot \dd x
=  - \frac{1}{2\pi} \dd n^\mu \, \partial_\mu  T_p(n) = - \frac{1}{2\pi}\dd  T_p(n) \,.
\label{eq:classicalEMlgt}
\]
so that the field is manifestly pure gauge. Note that the expectation of the quantum field is also purely angular: in comparing to classical expressions, we should bear in mind that the quantum field implicitly involves an angular projection.

The choice of dressing connects precisely to ambiguity in the spectral density as follows.
In equation~\eqref{eq:basicLadderCRs}, we introduced ladder operators $A(p)$ and $A^\dagger(p)$ for the massive states.
For the dressed states, we similarly introduce dressed ladder operators: 
\[
\DressedA{T}(p) = 
 e^{i \SoftCharge{T}} A(p) e^{-i \SoftCharge{T}}\,.
\]
It is straightforward to check that the dressed operators obey the usual algebra
\[
&[\DressedA{T}(p), \DressedA{T}(p')] = 0 \,, \\
&[\DressedA{T}(p), \DressedA{T}^\dagger(p')] = \delta_\Phi(p, p') \,.
\]
Further, the total momentum operator commutes with the annihilation operator as one would expect:
\[
&[\mathbb{P}^\mu_\text{tot}, \DressedA{T}(p)] = -p^\mu \, \DressedA{T}(p)\,.
\]
It is important here that the dressing involves only zero-energy photons in the idealised sense discussed above in section~\ref{sec:dressedState}.

Corresponding to the dressed ladder operators, we may define a dressed local quantum field by
\[
\DressedPhi{T}(x) &= e^{i \mathbb{P}_\text{tot} \cdot x} \int \dd\Phi(p) \big( \DressedA{T}(p) + \DressedA{T}^\dagger(p) \big) e^{-i \mathbb{P}_\text{tot} \cdot x} \\
&= \int \dd\Phi(p) \big( e^{-i p \cdot x} \DressedA{T}(p) + e^{i p \cdot x} \DressedA{T}^\dagger(p) \big) \,.
\]
The dressed propagator, in exact analogy with equation~\eqref{eq:KL} is then 
\[
i \Delta_F[T] (p) &= \int \dd^4 x \, e^{i p\cdot x} \braket{0 | \timeOrder \DressedPhi{T}(x) \DressedPhi{T}(0) | 0 } \\
&= \int_0^\infty \dd s \, \frac{\rho[T](s)}{p^2 - s + i \epsilon} \,.
\]
Upon using $Q_s |0\rangle =0 $ on account of the matter number operators, see Eq.~\eqref{eq:Qc}, the dressed spectral density is then
\[
\rho[T](s) &= \sum_n |\langle p, n|\phi[T](0)|0\rangle|^2 \delta(s-m_n^2) \\
&= \sum_n \left| \int \dd\Phi(\tilde p) \braket{p, n| 
e^{i \SoftCharge{T}} A^\dagger(\tilde p) | 0 } \right|^2 \delta(s - m_n^2) \,.
\]
Compare to the undressed case from equation~\eqref{eq:undressedSpectralDensity}:
\[
\rho(s) = \sum_n \left| \int \dd\Phi(\tilde p) \braket{p, n| A^\dagger(\tilde p) | 0 } \right|^2 \delta(s - m_n^2) \,.
\]
Evidently the dressing modifies the spectral density, as expected. 
Indeed, the modification is precisely through the soft charge, parametrised by precisely the shape function which also controls the large gauge transformation~\eqref{eq:classicalEMlgt}.

As mentioned, we choose the shape parameter $T_p(n)$ so that the expectation value~\eqref{eq:genericPhotonDressing} reproduces the Coulomb field of the charged particle. This field depends on the asymptotic (symmetry) frame and can be changed by large gauge transformations, which in this case form the electromagnetic asymptotic symmetry group. We therefore interpret such a dressed Fock state as describing the original particle in the asymptotic symmetry frame defined by $T_p(n)$ relative to the frame in which the asymptotic field vanishes identically.
We may equally well interpret the dressed state as a state in the original asymptotic (symmetry) frame, i.e. as a state describing the particle and some low-frequency radiation. 
Scattering of different such states sources spacetimes with different asymptotic properties~\cite{DeAngelis:2025vlf} at future null infinity, ${\cal I}^+$.
See also Refs.~\cite{Strominger:2013jfa, He:2014laa, Strominger:2014pwa, Strominger:2017zoo} for relations between asymptotic symmetry generators and the soft-particle content of asymptotic states.

At this point we are in a position to understand our choice~\eqref{eq:shapeAsDerivative} of $f_p^\helicity(n)$ in more detail.
We have made the gauge choice $U \cdot \epsilon^\helicity = 0$. 
In the rest frame of the observer, this means that the timelike component $\epsilon_0^\helicity$ of the polarisation vectors vanish.
Of course $n \cdot \epsilon^\helicity(n) = 0$ too, which in particular implies that $\hat{\v{x}} \cdot \v{\epsilon}^\helicity = 0$: the polarization vectors are purely tangential to the sphere whose normal is $\hat{\v{x}}$.
In this context it is useful to introduce the differential operator $\eth$, which acts on functions on the sphere, and is defined by
\[
\eth \coloneqq \epsilon^+(n) \cdot \partial \,.
\]
It is a fact that all spin-weight one functions on the sphere are obtained by acting with $\eth$ on ordinary functions on the sphere~\cite{Newman:1966ub, Goldberg:1966uu, Eastwood1982edth}. 
The spin weight is precisely the same as our helicity weight of the dressing functions, so the derivatives in our choice~\eqref{eq:shapeAsDerivative} are inevitable\footnote{We chose $T_p(n)$ to be real. More generally one could choose it to be complex and also consider different functions $T_p(n, \helicity)$ for the different helicities. We leave these interesting possibilities for future study.}.

In summary, we have extended the discussion in Ref.~\cite{Elkhidir:2024izo}, by connecting different choices of $T_p(n)$ to different choices for the asymptotic frame.
It would be interesting to understand the interpretation of more complicated dressings, for example involving squeezed states, etc.

\section{Dressings in gravity}
\label{sec:dressedGravity}

In this section, we begin by examining some first consequences of our dressings in the gravitational case.
We then discuss the gravitational fields of boosted black holes, emphasising that the angular components (``shear'') of the gravitational field vanishes if one boosts the Kerr-Schild form of the metric.
This allows us to determine the asymptotic frame appropriate for massive charged particles with no dressing.

\subsection{Dressing for a general BMS supertranslation}
\label{sec:dressedGravityBMS}

Compared to the electromagnetic case of the previous section, gravitational dressings differ only because the mediators have spin 2.
We can therefore follow the discussion of section~\ref{sec:basicSetup} closely.

The zero energy dressings of interest to us are generated by a soft charge
\[\label{eq:QsOfF}
\SoftCharge{f} =  \sum_\helicity\int\! \dd\Phi(k, p)\,
      A^\dagger(p) A(p) \, \hdelta (\omega) \bigl(f_p^\helicity(n) \,a_\helicity^\dagger(k)+\bar{f}_p{}_\helicity(n) \,a^\helicity(k)\bigr) \Big|_{k = \omega n}\,,
\]
where the zero-energy shape function $f^\helicity_p(n)$ was introduced in equation~\eqref{localization}.
In gravity, $f^\helicity_p(n)$ is a spin-weight 2 function on the sphere, so,
in contrast to the single derivative relevant for the electromagnetic case~\eqref{eq:shapeAsDerivative}, 
we obtain $f_\helicity$ in gravity by differentiating an ordinary function $T_p(n)$ on the sphere twice:
\[
f^\helicity_p(n) = -2\bar{\epsilon}^\eta{}^{\mu}(n) \bar{\epsilon}^\eta{}^{\nu}(n) \frac{\partial}{\partial n^\mu}
\frac{\partial}{\partial n^\nu}
T_p(n) \,.
\label{eq:GRf}
\]
The normalisation factor of $-2$ is convenient below. 
Notice that we have written the gravitational polarisation tensor as an outer product of two spin one polarisation vectors.
The corresponding soft charge, now taken to be defined by the choice of $T_p(n)$, is then 
\[
\SoftCharge{T} =  \frac{-4\pi}{\kappa}\sum_\helicity \int\! \dd\Phi(k, p)
      A^\dagger(p) A(p) &\hdelta (\omega) \bigl(\bar{\epsilon}^\helicity_\mu(n) \bar{\epsilon}^\helicity_\nu(n) \, \partial_n^\mu \partial_n^\nu \, T_p(n)\,a_\helicity^\dagger(k)\\ 
      &+\epsilon_\helicity^\mu(n)\epsilon_\helicity^\nu(n) \, \partial_n{}_\mu \partial_n{}_\nu \, T_p(n) \,a^\helicity(k)\bigr) \Big|_{k = \omega n}\,,
\label{eq:QcG}      
\]
where $\partial^\mu_n = \partial/\partial n_\mu$.
Correspondingly, a gravitationally dressed, localised, point particle is given by
\[\label{eq:gravDressing}
\DressedStateR{\psi}{T} \coloneqq \int \dd\Phi(p) \, \varphi(p) \, e^{i \SoftCharge{T}} |{p}\rangle 
= e^{i \SoftCharge{T}}
|{\psi}\rangle 
\ .
\]
Although our notation does not distinguish between electromagnetic and gravitational dressings, we hope that context will make our meaning clear.

Specialising the generic stationary phase expression~\eqref{eq:statPhase} to the gravitational case, the metric perturbation quantum field is\footnote{We choose to insert factors of $\kappa$ so that $\mathbb{h}$ is dimensionless and $T_p(n)$ has dimensions of length. 
As the Einstein equations with source lead to
asymptotic fields of order $G \sim \kappa^2$, the soft charge in Eq.~\eqref{eq:QcG} is of order~$\kappa$.}
\[\label{eq:gravitonStatPhase}
\mathbb{h}^{\mu\nu}(x) \simeq \sum_\helicity\frac{-i\, \kappa}{4\pi r} \int_0^\infty  \dd \omega\, \epsilon_\helicity{}^{\mu}(k)\epsilon_\helicity{}^{\nu}(k) a^\helicity(k) e^{-i \omega u} 
\big|_{k = \omega n} + \textrm{h.c.} +{\cal O}(1/r^2)\,.
\]
The expectation value of this operator on the dressed point particle state~\eqref{eq:gravDressing} is easily found to be
\begin{align}
\label{eq:coulomb}
h_{\mu\nu}(x) &\coloneqq 
\DressedStateL{\psi}{T}
\mathbb{h}_{\mu\nu}(x) 
\DressedStateR{\psi}{T}
\\
&= -\frac{1}{r} \sum_\helicity 
\big(\epsilon_\eta{}_\mu \epsilon_\eta{}_\nu \bar\epsilon^\helicity{}_\rho \bar\epsilon^\helicity{}_\sigma + \bar\epsilon^\helicity{}_\mu \bar\epsilon^\helicity{}_\nu \epsilon_\eta{}_\rho \epsilon_\eta{}_\sigma\big) \partial^\rho \partial^\sigma T_p(n) \,.
\end{align}
The helicity sum can be performed using the projector of equation~\eqref{eq:projectorDef}, with the familiar result
\[
\sum_\helicity 
\epsilon_\eta{}_\mu \epsilon_\eta{}_\nu \bar\epsilon^\helicity{}_\rho \bar\epsilon^\helicity{}_\sigma  =  \frac12 \left( P_{\mu \rho} P_{\nu \sigma} + P_{\nu \rho} P_{\mu \sigma} - P_{\mu\nu} P_{\rho \sigma} \right) \,.
\]

In view of the gauge conditions on our polarisation vectors, it is clear that (as in electrodynamics) we are only sensitive to the angular components of $h_{\mu\nu}(x)$.
In this respect we make contact with asymptotic expansions of classical spacetime metrics, which is the subject of the celebrated Bondi formalism.
The relevant angular components of the metric are known as the ``shear'' $C_{AB}$.
Readers unfamiliar with Bondi coordinates\footnote{
It is interesting to note that IR divergences raise questions regarding the assumption that Bondi coordinates exist for scattering spacetimes~\cite{DeAngelis:2025vlf}, depending on the dressing of the massive particles being scattered. 
Here we treat spacetime as perturbative, that is we assume that particles propagate on flat space and the corrections to spacetime at large but finite distances due to gravitational wave emission are included perturbatively.
}, supertranslations etc will find a brief review in Appendix~\ref{app:BMS}.

To extract the shear from Eq.~\eqref{eq:coulomb}, we pass from $h_{\mu\nu}$ to $h_{\mu\nu} \dd x^\mu \dd x^\nu$ using the explicit differentials~\eqref{eq:dcoords},
finding
\[
h_{\mu\nu} \dd x^\mu \dd x^\nu &= -2r \, \dd \theta^A \dd \theta^B \, \partial_A n^\mu \partial_B n^\nu \left( P_{\mu\rho} P_{\nu \sigma} - \tfrac12 P_{\mu\nu} P_{\rho\sigma} \right) \partial^\rho \partial^\sigma T_p(n)  \\
&= r\,  \dd \theta^A \dd \theta^B \left(  \Omega_{AB} D^2 -2 D_A D_B \right) T_p(n) \\
&= r \,  \dd \theta^A \dd \theta^B \, C_{AB} \,,
\]
where the $\theta^A$ are coordinates on the (round) celestial sphere $S^2$, and $D$ is the corresponding covariant derivative.
The shear is therefore 
\[
C_{AB} = \left(  \Omega_{AB} D^2 -2 D_A D_B \right) T_p(n) \,,
\label{eq:shear_shift_from_conjugation}
\]
as expected from Eq.~\eqref{supertranslation}.

Before moving on, let us make two small comments. 
First, notice that the differential operator appearing in the definition of $\SoftCharge{T}$
annihilates the $Y_{00}$ and $Y_{1m}$ spherical harmonics, and thus it projects them out of the supertranslation parameter $T_p(n)$. While in the following we will not always make this projection manifest, we will discard these terms when they can be exposed without leading to a cumbersome notation. 
Second, in light of the discussion in Ref.~\cite{Elkhidir:2024izo} and as noted in Sec.~\ref{sec:dressedState}, our dressings can be interpreted as a resummation of certain modified tree-level 3-point matter-mediator amplitudes.  
We note here that, at least in gravitational theories, no higher-order corrections to these 3-point amplitudes are required. Indeed, on dimensional grounds such amplitudes fall off faster than $1/r$ in position space, so they do not represent a change in the shear. They should therefore be identified with corrections to the ${\cal O}(1/r^{n\ge 2})$ terms in Eq.~\eqref{eq:coefexpansion}, which are also nonlinear in $\delta C_{AB}$ (under our current assumption that we start from a spacetime with no shear).

\subsection{The canonical frame from the Kerr-Schild metric}
\label{sec:VVKS}

As we have emphasised, expectation values of quantum fields capture angular components of the corresponding classical fields in electrodynamics and gravity.
This raises the question: in gravity, what metric perturbation corresponds to the ``canonical'' choice $A(p)$ of matter ladder operator? 
That is, what asymptotic frame, if any, does this basic choice in quantum field theory actually correspond to?
Our purpose now is to show that there is an appropriate asymptotic frame, which coincides (at large distances, and in linearised gravity) with the celebrated Kerr-Schild choice of coordinates.

Of course we are really interested in situations where at least two black holes interact.
In the far past, long before any scattering has taken place, the classical description linearises because the black holes are very distant from one another.
As a result, the spacetime metric is a sum of contributions from each black hole and so we can begin our classical discussion by considering the familiar linearised metric of a single black hole.

A first question is what linearised spacetime metric should we start with? 
Schwarzschild black holes are commonly discussed in a variety of coordinate systems, all of which are related by diffeomorphisms --- but in this situation we must be very careful about the asymptotic behaviour of these diffeomorphisms.

We start with the Schwarzschild black hole in Kerr-Schild coordinates, in which the line element is
\[\label{eq:KSlineElement}
\dd s^2 = \dd s_0^2 - \frac{2GM}{R} \dd (T-R)^2 \,.
\]
We have written the flat Minkowski line element as $\dd s_0^2$, writing $T$ and $R$ for the (Kerr-Schild) time and radius in a frame in which our black hole is static.
The Kerr-Schild form of the metric is particularly interesting for many reasons.
Although the expression~\eqref{eq:KSlineElement} is obviously linear in $GM$, this is nevertheless an exact solution of the full Einstein equations.
The Kerr-Schild form has an obvious double-copy structure which is the foundation~\cite{Monteiro:2014cda} of the classical double copy.
For our purposes at the moment, we regard the Kerr-Schild form as simply a solution of linearised gravity.

A second form of the linearised solution which will be important for us is the De Donder gauge metric given by
\[
g_{\mu\nu}(x) = \eta_{\mu\nu} + h_{\mu\nu}(x) \,,
\]
with
\[\label{eq:ddH}
h_{\mu\nu}(x) = -\frac{4GM}{R} \left( v_\mu v_\nu - \frac{1}{2} \eta_{\mu\nu} \right) \,.
\]
In this form, the (constant in the far past) proper velocity of the black hole is written as $v_\mu$.
The De Donder perturbation~\eqref{eq:ddH} has the important property
\[
\partial^\mu \bar{h}_{\mu\nu}(x) = 0\, , \qquad
\bar{h}_{\mu\nu} = h_{\mu\nu} - \tfrac12 \eta_{\mu\nu} h^\rho_{\, \rho} \,.
\]
This is very helpful because the linearised Einstein equations become the wave equation for $\bar{h}_{\mu\nu}$, sourced by the black hole energy-momentum tensor.
This makes the De Donder perturbation a particularly natural starting point for a perturbative solution of full Einstein equation, including in scattering situations.

Since both the Kerr-Schild~\eqref{eq:KSlineElement} and De Donder~\eqref{eq:ddH} metrics are solutions of the linearised Einstein equation with the same source, they must be gauge transformations of one another:
\[
g_{\mu\nu}^\text{KS}(x) = g_{\mu\nu}(x) + \nabla_\mu \delta x_\nu + \nabla_\nu \delta x_\mu \, ,
\]
where $g_{\mu\nu}^\text{KS}(x)$ is the Kerr-Schild metric and $\delta x_\mu$ is the linearised diffeomorphism.
It is not difficult to check that the relevant diffeomorphism is 
\[\label{eq:KSmeetsDD}
\delta x_\mu = 2GM \, v_\mu \log R + GM \partial_\mu R \,.
\]
We emphasise that this is a large diffeomorphism: it does not fall off at large distance $R$. (See Ref.~\cite{Vines:2017hyw} for more discussion of diffeomorphisms relating various forms of the Schwarzschild metric.)

Radiation of energy, momentum and angular momentum is understood in full general relativity in another set of coordinates: the Bondi coordinates.
In Bondi form, the line element is (see Appendix~\ref{app:BMS})
\[\label{eq:BondiForm}
\dd s^2 = g_{uu} \dd u^2 + 2 g_{ur} \dd u \, \dd r + 2 g_{uA} \dd u \, \dd x^A - r^2 \left(\Omega_{AB} +\frac{C_{AB}}{r} + \cdots \right) \dd \theta^A \, \dd \theta^B \,,
\]
where, as before, $\Omega_{AB}$ is the round sphere metric. We have written the coordinates on the two-sphere as $\theta^A$, where $A = 1,2$ as usual. The shear $C_{AB}$ is the subdominant term in the metric restricted to the two-sphere; in Bondi form, we require that the trace $\Omega^{AB} C_{AB} = 0$.
The question now is whether the Kerr-Schild or De Donder metrics are in Bondi form.

More precisely, we are really interested in whether the total linearised metric describing our scattering black holes is in Bondi form. 
As the black holes have different rest frames we must consider boosted versions of the Kerr-Schild and De Donder metrics in which each black hole has a different velocity.
A first step is to consider a single black hole, boosted to velocity $v^\mu$. 
We have already written the De Donder expression in an appropriate form~\eqref{eq:ddH}. 
For the Kerr-Schild case, we note that
\[\label{eq:KScoordsCartForm}
T &= v \cdot x \,,\\
R &= \big((v \cdot x)^2 - x^2\big)^{1/2} \,.
\]
To set up a Bondi version of these Kerr-Schild coordinates, we consider an observer at a point $x^\mu$ which is very far from the black hole and let $U^\mu$ be the proper velocity of this observer. We can again write the observer's coordinates as
\[\label{eq:BondiTypeCoords}
x^\mu = u U^\mu + r n^\mu \,,
 \]
where $r$ is the radial distance and $n^\mu$ is a null vector which depends on the angles $x^A$ and satisfies $n \cdot U = 1$.
To see why these coordinates are relevant, note that we can write the flat metric as
\[
\dd s_0^2 = \dd x \cdot \dd x &= \dd u^2 + 2 \, \dd u \, \dd r + r^2 \dd n \cdot \dd n \\
&= \dd u^2 + 2 \, \dd u \, \dd r - r^2  \Omega_{AB} \, \dd x^A \, \dd x^B\\
&= \dd u^2 + 2 \, \dd u \, \dd r +  e_n \cdot e_n \,.
\]
In the last line above, we defined the (two dimensional, spherical) one-form frame
\[
e_n^\mu \coloneqq r \, \dd n^\mu 
\]
which is evidently normalised to unity. 
This choice of frame is helpful because it simplifies the large $r$ asymptotics, and hence the extraction of the shear.

Our choice of coordinates brings the flat space metric into Bondi form but, of course, we are more interested in the curved metric. Continuing with the Kerr-Schild form, by expanding $T$ and $R$ in equation~\eqref{eq:KScoordsCartForm} at large $r$ we find that
\[
\dd (T-R) = \frac{\dd u }{v \cdot n} + O(1/r) \,.
\]
Hence the line element is
\[
\dd s^2 = \dd u^2 + 2 \dd u \, \dd r + e_n \cdot e_n - \left( \frac{2 GM}{r (v\cdot n)^3} \dd u^2 + \mathcal{O}(1 / r^2 ) \right) \,.
\]
This is directly in the desired Bondi form, with shear explicitly zero, $C_{AB}=0$. 
With a little more work one can expand the metric to any order in $1/r$; as observed in reference~\cite{Bonga:2018gzr} it retains the Bondi form to a surprisingly high order in $1/r$.

By performing a large gauge transformation an asymptotically-simple metric can always~\cite{Flanagan:2015pxa}  be put in Bondi form with vanishing shear on $\mathcal{I}^-$.
This particular choice is known as the \emph{canonical} frame.
Thus we have demonstrated that a boosted black hole in Kerr-Schild coordinates is canonical~\cite{Bonga:2018gzr}.
The fact that the Kerr-Schild form is canonical persists when we consider several black holes of masses $M_i$ and velocities $v_i$. Then at large distances the metric reads
\[
\dd s^2 = \dd u^2 + 2 \dd u \, \dd r + e_n \cdot e_n - \left( \sum_i \frac{2 GM_i}{r (v_i\cdot n)^3} \dd u^2 + \mathcal{O}(1 / r^2 ) \right) \,.
\]

Now we return to the De Donder form of a single boosted black hole, asking whether this is also in the canonical frame, or instead is in a different frame?
At large radius $r$, the De Donder line element becomes
\[\label{eq:ddNotBondi}
\dd s_0^2 - &\frac{4GM}{R} \big( v_\mu v_\nu - \frac12 \eta_{\mu\nu} \big) \dd x^\mu \, \dd x^\nu \\
&= \dd s_0^2 - \frac{4GM}{r \, n \cdot v} \left( ( v \cdot U \, \dd u + n \cdot v \, \dd r + v \cdot e_n)^2 - \frac12 \dd u^2 - \dd u \, \dd r - \frac 12 e_n \cdot e_n \right) \,,
\]
up to corrections of order $1/r^2$.
This is clearly not in Bondi form: for example, unlike equation~\eqref{eq:BondiForm} the coefficient of the $\dd r \, \dd r$ term in the line element~\eqref{eq:ddNotBondi} has a non-vanishing coefficient given by
\[
g_{rr}^\text{DD} = - \frac{4GM}{r} n \cdot v \,.
\]
The situation actually simplifies somewhat if we consider scattering several black holes, choosing our observer to be at rest in the associated CM frame. Then it follows that
\[
\sum_i M_i v_i^\mu = E U^\mu \,,
\]
where $E$ is the total CM energy.
The $\dd r \, \dd r$ term in the line element is now given by
\[
g_{rr}^\text{I} = - \frac{4GE}{r} \,,
\]
and the mixed $g_{rA}$ term vanishes because $U \cdot n = 1 \Rightarrow U \cdot e_n =0$.
Further, we can remove the unwanted $g_{rr}^\text{I}$ term in the metric by a redefinition of the $u$ coordinate:
\[\label{eq:intrinsicShift}
u \rightarrow u + 2 GE \log r \,.
\]
This is a large diffeomorphism, but crucially it does not depend on angles.
It represents a shift of the origin of the time coordinate for all observers at a fixed distance.
This redefinition of $u$ is independent of angles, so it does not affect the shear. 
However the shear as it stands is not traceless; enforcing this requirement involves an additional redefinition of the radius by a constant $\delta r \propto GE$.
The combination brings the De Donder line element into Bondi form with shear: 
\[\label{eq:intrinsicShear}
C_{AB} = -4G \sum_i \frac{M_i}{n \cdot v_i} \left(v_i \cdot e_A v_i \cdot e_B - \frac12 \Omega^{CD} (v_i \cdot e_C)(v_i \cdot e_D)\Omega_{AB} \right)\,,
\]
where $e_n^\mu = r \, e_A^\mu \, \dd \theta^A$. 
We refer to the (Bondi) form of the De Donder metric, after the shift~\eqref{eq:intrinsicShift} as the \emph{intrinsic} frame, following Veneziano and Vilkovisky~\cite{Veneziano:2022zwh}.

Our understanding is that the intrinsic frame is the one relevant for MPM-PN computations, see e.g. Refs.~\cite{Blanchet:1985sp, Blanchet:1989ki, Blanchet:2013haa}.
Scattering amplitude computations, assuming an initial choice of ladder operator with $\braket{\psi | a_h(k) | \psi} = 0$, are evidently set up in the canonical frame.
The diffeomorphism between these is just the relation~\eqref{eq:KSmeetsDD} between the Kerr-Schild and De Donder frames.
The most important part of this diffeomorphism is the shift of the retarded time $\delta u$. Noting from equation~\eqref{eq:BondiTypeCoords} that $u = x \cdot n$ this shift is, in the case of several black holes,
\[
\delta u = n \cdot \delta x = \sum_i  \big( 2 G M_i \, n\cdot v_i \log n \cdot v_i  +  G n\cdot v_i  \, Y \big) \,,
\]
where $Y$ is independent of angles.
The first term on the right-hand-side above is the Veneziano-Vilkovisky supertranslation. The second term, which like~\eqref{eq:intrinsicShift} is independent of angles in the COM, can be interpreted as a change of the origin of the proper time and is therefore less significant.

%%%%%%%%%%%%%%%%%%%%%%%%%%%%%%%%%%%%%%%%%%
\section{The exact hard charge for general dressing}
\label{sec:general_hard_charges}

In the previous two sections, we have seen that the action of dressing massive particles has the classical effect of performing a large symmetry transformation.
This suggests that it may be possible to complete the soft charge $Q_s$ into a conserved charge $Q$ which commutes with the $S$-matrix~:
\[
[\Smatrix, Q] \equiv [\Smatrix, Q_s+Q_h]=0 \ ,
\]
see e.g. Ref.~\cite{Strominger:2017zoo}.
We will later see that, in fact, this also generalises to theories with no local symmetries.

Noticing that $Q_s$ is linear in creation and annihilation operators, the commutator is sensitive only to a single asymptotic mediator state at a time, and receives distinct contributions from that soft mediator being attached to a massless or massive particle,
\[
[\Smatrix, Q_s] = C_0+C_M \ ,
\label{eq:C0andCM}
\]
respectively; the former exists only in theories with self-interacting mediators. Equating (the negative of) this commutator with $[\Smatrix, Q_h]$ determines $Q_h$, though not uniquely as it can be shifted arbitrarily by other symmetries of the $S$ matrix. We will refer to the contribution compensating $C_0$ as the ``massless hard charge'' and denote it by $Q_{h, 0}$, and to the contribution compensating $C_M$ as the ``massive hard charge'' and denote it by $Q_{h, M}$.
These components were constructed from a general relativity perspective in Ref.~\cite{Campiglia:2015kxa}.
As we will see shortly, $Q_h$ effectively rescales $S$-matrix elements in ways that correlate pairs of external particles. 
The existence of a conserved charge $Q=Q_s+Q_h$ indicates quantitatively that soft/zero-energy dressing of asymptotic states is equivalent to a suitable {\em finite} rescaling of $S$-matrix elements for bare asymptotic states. 

Except for Sec.~\ref{sec:generalstrategy} which lays out the general strategy for determining $Q_h$, in the rest of this section we describe this calculation for a general dressing in a gravitational theory.
The calculations go through with minor technical adjustments for vector and massless scalar mediators. 
In Sec.~\ref{sec:VV} we will further focus on the dressing connecting the canonical and intrinsic BMS frames, and discuss the consequences of finite transformations. Both here and in Sec.~\ref{sec:VV} we will keep the complete dependence on the exchanged momentum. We will then see in Sec.~\ref{sec:consequences} that the connection between the asymptotic charges and the multi-particle states involves effectively a(n eikonal) resummation. This effectively identifies the exchanged momentum with the impulse, which needs not be ${\cal O}(\hbar)$.

\subsection{General strategy for static dressing}
\label{sec:generalstrategy}

The commutator of the GR soft charge in Eq.~\eqref{eq:QsOfF} and the $S$ matrix is 
\begin{align}
\label{eq:GendefCommutator}
C \equiv \frac{2\pi}{\kappa}\sum_\helicity \int \dd\Phi(k,p) \hdelta(\omega)  f{}^\helicity_p(k)\, [\Smatrix, A^\dagger(p) A(p)\,a^\dagger_\helicity(k)] \,. 
\end{align}
Because of the effectively zero frequency of the mediators in the dressing, only the leading soft limit of the $S$ matrix is relevant. That is, we may write 
\begin{align}
%\hspace{-8pt}
\Smatrix_L &= -
%i
\sum_\helicity  \int \dd\Phi(k, P, P')  
\frac{
%\mathcal{A}_3(P, k_\helicity  \rightarrow P+k)
 (i H(P, P')) \SoftTO{P}{k_\helicity}{P+k}}{2 P\cdot k + i\varepsilon}  a_\helicity (k) \,,
\\
%\hspace{-5pt}
\Smatrix_R &= -
\sum_\helicity \! \int \!\dd\Phi(k, P, P') \frac{
%\mathcal{A}_3(P'-k, k_\helicity  \rightarrow P')
 (i H(P, P')) \SoftTO{P'-k}{k_\helicity}{P'}}{-2 P'\cdot k + i\varepsilon}  a_\helicity (k) \ ,
\end{align}
for the two orientations of $k$ relative to outgoing momenta $P$ and $P'$.
Here ${\cal A}$ is the soft factor due to the soft momentum $k$,
\[
%\mathcal{A}_3(p+k \rightarrow p, k_\helicity) 
\SoftTO{P}{k_\helicity}{P+k} = \kappa {\bar \epsilon}{}^{\helicity}_{\mu\nu} P^\mu P^\nu \equiv \kappa {\bar {\cal A}}{}^+_\helicity
~,\qquad
\SoftTO{P'-k}{k_\helicity}{P'} = \kappa {\bar \epsilon}{}^{\helicity}_{\mu\nu} P'{}^\mu P'{}^\nu \equiv \kappa {\bar {\cal A}}{}^-_\helicity
\,,
\label{eq:notationA+A-}
\]
with ${\bar \epsilon}{}^{\helicity}_{\mu\nu}={\bar \epsilon}{}^{\helicity}_\mu{\bar \epsilon}{}^{\helicity}_\nu$
and $H(P,P')$ refers to the generic ``hard'' matrix element with all external particles other than the soft mediator, and we have explicitly written the annihilation operator of the soft particle.
These relations hold for both massive and massless $P$ and $P'$. While in the classical limit we always have pairs of particles of the same mass, we may have unpaired massless particles.   

In terms of the $\kappa$-less soft factors defined in Eq.~\eqref{eq:notationA+A-} the explicit $\kappa$ dependence of the commutators cancels out and they become
\[
C^+ &= -{2\pi}
\sum_\eta \int \dd\Phi(P, P') (i H(P, P'))
 \int \dd \Phi(l) \hdelta(\omega_l) \,  
 \frac{{\bar {\cal A}}{}^{+}_\eta\, f^\eta_p(l)}{2P\cdot l + i\varepsilon} \ ,
 \\
 C^- &= -{2\pi}
 \sum_\eta \int \dd\Phi(P, P') (i H(P, P'))
 \int \dd \Phi(l) \hdelta(\omega_l) \, 
 \frac{{\bar {\cal A}}{}^{-}_\eta\, f^\eta_p(l)}{-2P'\cdot l + i\varepsilon} \ . 
\]
The delta function from the commutator of the creation and annihilation operators allowed us to evaluate the integral in the definition of $Q_s$;
the frequency factor from that commutator combines with the factor of $\omega_l^{-1}$ in the phase space measure $\dd\Phi(l)$ to yield a finite value for the remaining integral in the presence of $\hdelta(\omega_l)$.

Since the $l$ integral depends only on the momentum $p$ of the particle carrying the soft charge and the momentum $P$ or $P'$ to which the low-energy mediator is attached, these commutators can be cancelled by an operator proportional to the number operator for the particles whose interactions are governed by the hard $S$-matrix element $H(P, P')$.
We note that this mechanism appears to hold generically, even in the absence of known asymptotic symmetries and, in a sense it supports the idea that amplitude soft factorization in the soft limit \emph{implies} the existence of a conserved charge. We will briefly return to this observation in the concluding section, but otherwise we will not pursue its consequences.

The $l$ integral can be computed on a case by case basis, given the explicit expressions for the soft factors ${\cal A}^\pm$. We will outline below this calculation for a graviton mediator, for which the soft factor is the stress tensor of the particle not carrying the soft charge.

To regularize the integrals we will use a variant of the $\beta$-regulator of Refs.~\cite{Bini:2024rsy, Elkhidir:2024izo}, which pushes the momentum of the particle carrying the soft charge slightly off-shell, and thus moves the support of the $k$ integral away from the origin. 
Because of the more general structure of the integral, the direct connection between the off-shellness of the particle carrying the soft charge and the graviton energy is no longer apparent, so we will simply replace uniformly
\[
\hdelta(\omega) \mapsto \hdelta_\beta(\omega) = \hdelta (\omega-\beta) \ .
\]
If an additional regulator is required, we will use dimensional regularization, and the order in which the regulators are removed is that we \emph{first} take $d\rightarrow 4$ and \emph{afterwards} $\beta\rightarrow 0$.

\subsection{Hard massless charge}
\label{sec:hard_massless_general}

It is convenient to expose the uniformity of the massive and massless contributions and define the integral
\[
\label{eq:masslesstensorintegral}
I^p_0(K)_{\mu\nu} = \sum_\eta \int \dd \Phi(l) \hdelta_\beta(\omega_l) \frac{{\bar \epsilon}(l)^\eta_{\mu \nu} \epsilon(l)_\eta^{AB} (\Omega_{AB} D^2 -2 D_A D_B) T_p(n_l)}{2 K \cdot l + i \epsilon} \ .
\]
Here the superscript $p$ is the momentum of the particle carrying the soft charge and the $0$ subscript emphasizes that the momentum $K$ is massless, $K^2=0$, which is the focus of this section. In the next section we will discuss massive momenta, and drop the $0$ subscript.
%%%
The on-shell condition in the phase space integral sets $l^\mu = \omega_l n_l^\mu(\theta, \phi)$; we also use the notation $\epsilon(l)_\helicity^{AB} = \epsilon_\helicity^{\mu\nu} e_\mu^Ae_\nu^B = \epsilon(l)_\helicity^{\mu\nu} D^A n_l{}_\mu D^B n_l{}_\nu$ and  $P^A = P^\mu e_\mu^A = P^\mu D^A n_l{}_\mu$ with $D^A$ the covariant derivative on the celestial sphere.
With this definition, the contribution of a massless external graviton to the commutator of the $S$ matrix and the soft charge, $C_0$ in Eq.~\eqref{eq:C0andCM}, is
\begin{align}
%\label{eq:Cmassive1Dgen}
C_0 =&  \int \dd\Phi(P',P)\, (i H(P,P'))  \left[\;  - T^{\mu\nu}(K) I^p_0(K){}_{\mu\nu}\right]
\end{align}
where $T^{\mu\nu}(K) = K^\mu K^\nu$ is the graviton stress tensor, and the square parenthesis is the relevant contribution to the massless hard charge.

The scalar integral governing this commutator, 
\[
I^0_{p, K} \equiv K^\mu K^\nu I^p_0(K)_{\mu\nu} &= \sum_\eta \int \dd \Phi(l) \hdelta_\beta(\omega_l) \frac{K^\mu K^\nu \epsilon(l)^\eta_{\mu \nu} \epsilon(l)_\eta^{AB} (\Omega_{AB} D^2 -2 D_A D_B) T_p(n_l)}{2 K \cdot l + i \epsilon} \\
&= \int \dd \Phi(l) \hdelta_\beta(\omega_l) \frac{ K^A K^B \, (\Omega_{AB} D^2 -2 D_A D_B) T_p(n_l)}{2 K \cdot l + i \epsilon} \ ,
\]
is evaluated\footnote{Here, $K$ is a 4-vector. Its relation to the vector in Appendix~\ref{app:int_general_dressing} is ${\hat K} = K/\omega_K=(1, \Kvec)=(1, \Khat)$ and $\Kvec = \Khat$ because for a massless vector $\Kvec^2= \nK^2 = 1$.}  in Appendix~\ref{app:int_general_dressing}, see Eq.~\eqref{eq:integral_final}.
The result, accounting for the fact that $(\Omega_{AB} D^2 -2 D_A D_B)$ annihilates $\ell=0$ and $\ell=1$ modes, is
\[
I^0_{p, K} = \frac{-1}{2\pi} \,\omega_K\,T^{\ell\geq 2}_p({\hat K}) \ ,
\label{eq:IK}
\] 
where the notation $T^{\ell\geq 2}_p({\hat K})$ indicates that the $\ell=0$ and $\ell=1$ modes are projected out of the original supertranslation parameter $T_p({\hat K})$.
The importance of this projection in the comparison of the MPM and QFT approaches to the gravitational waveform was recently emphasized in Ref.~\cite{Bini:2026dvn} through 3.5PN orders; Eq.~\eqref{eq:IK} shows that, as conjectured there, this projection is needed to all PN orders.

Putting this all together, the hard massless charge, chosen so that it cancels the massless external leg contribution $C_0$ to the commutator of the soft charge and the $S$ matrix, is
\[
Q_{h, 0} &=  \sum_\helicity \int \dd \Phi(k) \, Q^0_{h}(k) \, a^\dagger_\helicity (k) a_\helicity (k)
%\\
\quad, \qquad
Q^0_{h}(k) =  2\pi\sum_i \, I^0_{p_i, k}
%\left[-I^0_{p, k}\right] 
\ ,
\label{eq:finalmassless}
\]
where the sum runs over all the external matter particles carrying the soft dressing.

Before proceeding to discuss the hard matter charge, it is worth pausing for a moment to discuss how the massless hard charge we just derived yields the remaining part of the supertranslation: the shift of the retarded time by the supertranslation parameter, see Eqs.~\eqref{eq:u_shift} and~\eqref{eq:shear_shift_from_conjugation}.
These two aspects really come together, as we review in Appendix~\ref{app:BMS} from a geometric perspective, see especially Eq.~\eqref{supertranslation}. 
Upon completing $Q_s$ to the complete charge $Q$ and pushing it past the $S$ matrix~\cite{Elkhidir:2024izo}, the observable whose final state expectation value we want to compute is conjugated as ${\cal O}\mapsto e^{-i (Q_s+Q_{h})} {\cal O} e^{i (Q_{h}+Q_s)} $. As we discussed in Sec.~\ref{sec:dressedGravityBMS}, for ${\cal O}=\mathbb{h}_{\mu\nu}$ conjugation by $Q_s$ yields the Coulomb field. Conjugation by $Q_h$ yields the remaining part of the supertranslation --- the shift of the retarded time. Indeed, 
\[\label{eq:HardChargePhase}
e^{-i Q_{h}} a_\helicity (k) e^{i Q_{h}} = e^{-i \omega_k \, 
T_p(n)} a_\helicity (k) \,,
\]
which leads to
\[
u\mapsto u+ T_p(n)
\]
upon plugging Eq.~\eqref{eq:HardChargePhase} in Eq.~\eqref{eq:gravitonStatPhase}.

\subsection{Hard matter charges}
\label{sec:hard_matter_general}

\begin{figure}
    \centering
        \includegraphics[width=0.9\linewidth]{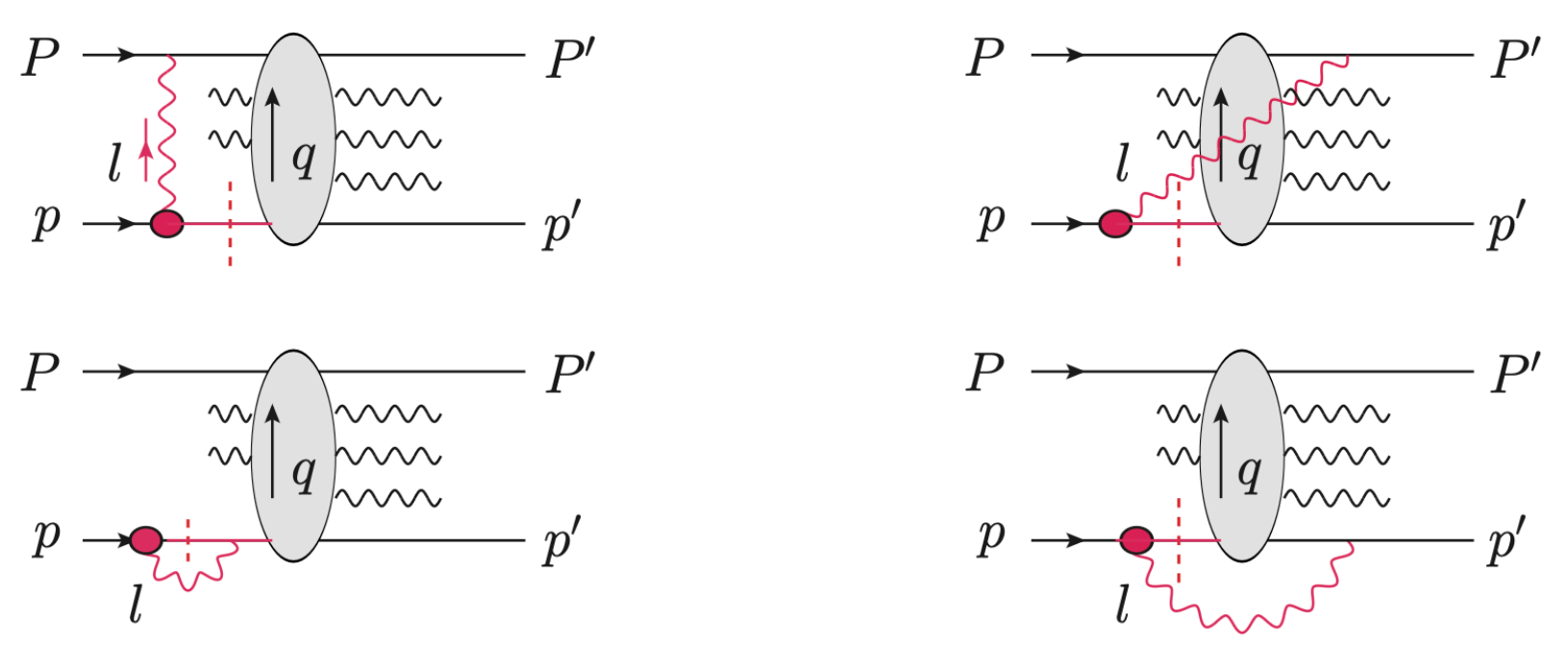}
    \caption{The diagrams contributing to the matter part of the commutator $[\Smatrix, Q_s]$ for a process with 4 external matter lines. The line crossed by the dashed line is cut; the (red) wiggly line denotes a cut mediator propagator. The 3-point blob on a matter line indicates the location of the dressing. The solid red cut line represents the factor of $\delta(\omega)$. In specific cases, such as the one discussed in Sec.~\ref{sec:VV}, it is an actual cut eikonal propagator.}
    \label{fig:4matter_general}
\end{figure}

The calculation of the hard matter charge for a time-independent soft dressing, compensating  $C_M$ in Eq.~\eqref{eq:C0andCM}, is similar to that of the massless hard charge. The main differences involve (1) focusing on pairs of contributions identified as corresponding to the same incoming and outgoing massive particle and (2) the details of the required integrals, i.e. the massive analog of \eqref{eq:masslesstensorintegral}, which now involve a massive momentum $P$ in place of the massless momentum~$K$. 
In addition, there are contributions in which the low-energy mediator starts and ends on the same matter particle, the second row in Fig.~\ref{fig:4matter_general}. We discuss these separately.

To parallel the setup of the massless hard charge, we define the tensor integrals 
\begin{align}
I_p(P){}_{\mu\nu} & = \sum_\eta \int \dd \Phi(l) \hdelta_\beta(\omega_l) \frac{{\bar \epsilon}(l)^\eta_{\mu \nu} \epsilon(l)_\eta^{AB} (\Omega_{AB} D^2 -2 D_A D_B) T_p(n_l)}{2 P \cdot l + i \epsilon} 
\\
J_p(p){}_{\mu\nu} & = \lim_{\beta\rightarrow 0} 
\frac{1}{p^2-m^2} 
\sum_\eta \int \dd \Phi(l) \hdelta_\beta(\omega_l) 
{{\bar \epsilon}(l)^\eta_{\mu \nu} \epsilon(l)_\eta^{AB} (\Omega_{AB} D^2 -2 D_A D_B) T_p(n_l)} \Big|_{\tiny{\begin{array}{l}{\omega\rightarrow\omega+\beta}\\{p\rightarrow p+\beta Y}\end{array}}} \ ,
\nonumber
\end{align}
where the vector $Y$ depends on the details of $T_p(n)$. 
As in the massless case, the on-shell condition in the phase space integral sets $l^\mu = \omega_l n_l^\mu(\theta, \phi)$ and we use the notation $\epsilon(l)_\helicity^{AB} = \epsilon_\helicity^{\mu\nu} e_\mu^Ae_\nu^B = \epsilon(l)_\helicity^{\mu\nu} D^A n_l{}_\mu D^B n_l{}_\nu$ and  $P^A = P^\mu e_\mu^A = P^\mu D^A n_l{}_\mu$ with $D^A$ the covariant derivative on the celestial sphere.

With these definitions, the contribution to the commutator $[\Smatrix, Q_s]$ from
an incoming-outgoing pair of massive particles is
\begin{align}
\label{eq:CmassiveFullgen}
C_M =& 
\int \dd\Phi(P',P)\, (i H(P,P'))  
\left[\; - T^{\mu\nu}(P') I_p(-P')_{\mu\nu} - T^{\mu\nu}(P) I_p(P)_{\mu\nu}\right]
\end{align}
and that from the same particle carrying $Q_s$, i.e. $p'{}^2=p{}^2=m^2$, is
\begin{align}
\label{eq:Cmassive1Dgen}
C_M^{\text{diag.}} =&  \int \dd\Phi(P',P)\, (i H(P,P'))  \left[\; 
- T^{\mu\nu}(p') I_p(-p'){}_{\mu\nu} - T^{\mu\nu}(p) J_p(p){}_{\mu\nu}\right]
\ .
\end{align}
where $T^{\mu\nu}$ is stress tensor for the corresponding matter particle.
The tracelessness of $I$ and $J$ integrals, actively imposed by the polarization factor $\epsilon_{\mu\nu}^\eta$, makes these stress tensors effectively be just
\[
T(P)^{\mu\nu} := P^\mu P^\nu 
\]
and the same for all other arguments, i.e. without the trace part proportional to $\eta^{\mu\nu}$.

The relevant integral for the evaluation of $C_M$ is
\begin{align}
I_{p,P} \equiv P^\mu P^\nu I_p(P)_{\mu\nu} &= \sum_\eta \int \dd \Phi(l) \hdelta_\beta(\omega_l) \frac{P^\mu P^\nu {\bar \epsilon}(l)^\eta_{\mu \nu} \epsilon(l)_\eta^{AB} (\Omega_{AB} D^2 -2 D_A D_B) T_p(n_l)}{2 P \cdot l + i \epsilon} \nonumber\\
&= \int \dd \Phi(l) \hdelta_\beta(\omega_l) \frac{ P^A P^B \, (\Omega_{AB} D^2 -2 D_A D_B) T_p(n_l)}{2 P \cdot l + i \epsilon} \ ,
\label{eq:massive_ini}
\end{align}
and it is evaluated in Appendix~\ref{app:int_general_dressing}, see Eq.~\eqref{eq:integral_final} with $\nK < 1$.
The result is~\footnote{Here, $P$ is a 4-vector. Its relation to the vector in Appendix~\ref{app:int_general_dressing} is $P = E_P (1, \Kvec)$ with $\Kvec^2\equiv \nK^2 <1$.}
\[
I_{p,P} &=  \frac{-1}{2\pi}(P^2)^2 \int_{\Sphere } 
\frac{{d^2\Omega}_{\bm n}}{4\pi} \, \frac{T^{\ell\ge 2}_p(n_l) }{(P\cdot n_l)^3} \ ,
 \label{eq:IP}
\]
where $n_l=(1, \nhat)$ with $\nhat^2 = 1$.
As discussed in the Appendix, this integral is finite because $P\cdot n \ne 0$ as $P$ is time-like and $n$ is null. As already discussed, the notation $T^{\ell\ge 2}_p(n_l) $ indicates that the $\ell=0, 1$ modes of $T_p(n_l) $ are dropped because they are projected out in the initial integral \eqref{eq:massive_ini} by the operator $(\Omega_{AB}D^2-2D_A D_B)$.

Eq.~\eqref{eq:IP} covers, up to simple relabeling, $I_{p,P}$, $I_{p,-P'}=-I_{p,P'}$ and $I_{p,-p'} = -I_{p,p'}$, the latter appearing in the diagonal contributions. 
The second diagonal contribution, given by the $J$ integral, is substantially simpler. Direct evaluation of the tensor integral using the delta functions yields
\[
J_p(p){}_{\mu\nu} & = -\frac{1}{2E_p}\lim_{\omega\rightarrow 0}
\sum_\eta \int \dd{}^2 \Omega_{\bm n} \; 
{{\bar\epsilon}(l)^\eta_{\mu \nu} \epsilon(l)_\eta^{AB} (\Omega_{AB} D^2 -2 D_A D_B) T_p(n_l)} \ .
\label{eq:Jp}
\]
The relevant diagonal integral is then
\[
J_p \equiv p^\mu p^\nu J_p(p)_{\mu\nu} \ ;
\]
we can evaluate it using the ingredients in Appendix~\ref{app:int_general_dressing}. The result is
\[
J_p \equiv p^\mu p^\nu J_p(p)_{\mu\nu} 
= \frac{6}{\pi E_p}\, \int 
\frac{d^2 \Omega_{\bm n}}{4\pi} \, (p\cdot n_l)^2 T^{\ell\ge 2}_p(n_l)
\label{eq:Jpp}
\ ,
\]
where, as discussed before, $ T^{\ell\ge 2}_p(n_l)$ indicates that we dropped terms containing only $Y_{00}$ and $Y_{1m}$ components of $T_p(n)$ which are already projected out in Eq.~\eqref{eq:Jp}. For the same reason, Eq.~\eqref{eq:Jpp} picks out only the $Y_{2m}$ parts of $T_p(n)$; we however do not manifestly enforce this in order to have a manifestly-covariant expression.

Combining Eqs.~\eqref{eq:IP}, its suitable relabelings, and \eqref{eq:Jpp}, the hard matter charge, chosen so that it cancels the massive external leg contribution $C_M$ to the commutator of the soft charge and the $S$ matrix, is then
\begin{equation}
\label{eq:Qhardmatter}
Q_{h, M} = \sum_i \int \dd\Phi(p_i)\,  Q^M_{h, i} \, A^\dagger_i A_i(p_i) \ ,
\end{equation}
where the sum is itself over all the types of matter particles\footnote{In the classical limit there are two particles of each type one incoming one outgoing, and the number operator in $Q_h$ will fire when it hits both of them.\label{foot:footnote9}}. $Q^M_{h, i}$ is a sum over all the incoming matter particles (because the ultra-soft graviton with momentum $k$ can attach to all matter particles and in $C_M$ we looked at an incoming particle and its outgoing counterpart). It is~\footnote{This can also be written in terms of the number operators of external matter particles.}
\begin{align}
Q^M_{h, i} =  2\pi\left[
%- 
I_{p_i, -p'_i} 
%- 
+J_{p_i} + \sum_{j\ne i}  \left[
%-
I_{p_i, -p'_j} 
%- 
+I_{p_i, p_j} \right] \right]\ .
\label{eq:finalmassive}
\end{align}
We may also interpret each term as a 2-particle hard matter charge contribution, corresponding to the two particles connected by the soft graviton. We will adopt this interpretation in Sec.~\ref{sec:VV}.

We emphasize that Eqs.~\eqref{eq:finalmassless} and \eqref{eq:Qhardmatter}-\eqref{eq:finalmassive}, together with the closed-form evaluation of the master integrals in Eqs.~\eqref{eq:IK}, \eqref{eq:IP}, and \eqref{eq:Jpp}, determine the hard charge $Q_h$ exactly, as a functional of the supertranslation parameter $T_p(n)$. No expansion in the momentum transfer, in the coupling, or in any other small parameter has been performed: the dependence of $Q_h$  on the external momenta enters only through the exact integrals $I_{p,P}$, $I_{p,-P'}$, $I_{p,p'}$, and $J_p$, each of which is a finite functional of $T_p(n)$ for any time-like $P$, $P'$ and any null $K$. In particular, the action of $Q_h$  on a multi-particle state is known completely once $T_p(n)$ is specified.

Before turning to applications we note here that, as also mentioned in Ref.~\cite{Elkhidir:2024izo}, the commutator of the hard and soft charges vanish, 
\[
[Q_s, Q_h] = 0 \ .
\]
Indeed, 
the soft charge $Q_s$ in Eqs.~\eqref{eq:QsOfF}-\eqref{eq:GRf} is linear in the creation and annihilation operators of soft mediators, while the hard charges $Q_{h,0}$ in Eq.~\eqref{eq:finalmassless} and $Q_{h,M}$ in Eqs.~\eqref{eq:Qhardmatter}-\eqref{eq:finalmassive} are built from number operators of mediators and of matter species, respectively. 
Number operators of one species commute with creation/annihilation operators of any other species, so $Q_s$ commutes manifestly with the matter part $Q_{h,M}$. For the mediator part $Q_{h,0}$, the $c$-number coefficient $Q^0_h(k)$ multiplying the mediator number operator vanishes linearly as $\omega \rightarrow 0$, see Eq.~\eqref{eq:IK}, so its commutator with the $\omega=0$ -supported operators in $Q_s$ likewise vanishes. 
We will use this property extensively in the next section, when detailing the time evolution of the coherent-state-dressed initial state.

%%%%%%%%%%%%%%%%%%%%%%%%%%%%%%%%%%%%%%%%%%
\section{The Eikonal final state and BMS dependence of impact parameter}
\label{sec:consequences}

The angular momentum of the gravitational radiation field depends on the asymptotic frame, see e.g.~\cite{Bonga:2018gzr,Compere:2019gft,Compere:2023qoa}. Since conservation of total angular momentum holds in all frames, it follows that the mechanical angular momentum of the matter particles should also depend on the asymptotic frame.
A simple interpretation, at least to leading order, is that the change in (the norm of the) orbital angular momentum is captured by a change in the impact parameter as measured at infinity,
\[
\delta J = p \, \delta b \ ,
\label{eq:extraJloss}
\]
see also Sec.~VI~C of Ref.~\cite{Bini:2022wrq}.
A similar interpretation, that the impact parameter depends on the BMS frame, was also advocated in Ref.~\cite{Compere:2023qoa}\footnote{See Sec.~9.4 in that reference.}.

In this section, we will compute the effective change of the impact parameter, induced by a general BMS supertranslation dressing, in a two-particle scattering event.
It is enough for us to consider purely conservative scattering because this is the setting where angular momentum has received most attention in the literature to date.
Because the hard charge $Q_h$ we constructed in Sec.~\ref{sec:general_hard_charges} is an exact functional of the dressing data $T_p(n)$, see Eqs.~\eqref{eq:finalmassless} and \eqref{eq:Qhardmatter}-\eqref{eq:finalmassive}, the resulting shift of the impact parameter is likewise exact. That is, it holds to all orders in perturbation theory, with the entire perturbative content carried by the $S$-matrix (restricted to conservative dynamics, and resummed as we discuss below) and by the kinematic dependence of the integrals discussed in the previous section.
In Sec.~\ref{sec:VV} we will apply this general result to the particular supertranslation connecting the canonical and intrinsic frames, see Sec.~\ref{sec:VVKS} and Ref.~\cite{Veneziano:2022zwh}, in the cases of scalar, vector and graviton mediators. 
We will first compute the frame dependence of the mechanical angular momentum to all orders, and then verify at leading order that Eq.~\eqref{eq:extraJloss} reproduces the known difference between the angular-momentum losses in the two frames, in GR~\cite{Bini:2022wrq} and in QED~\cite{Saketh:2021sri}.

Starting from a dressed initial state $\DressedStateR{\psi}{T}$, the expectation value of an operator ${\cal O}$ in the corresponding final state is 
\[
O 
= \DressedStateL{\psi}{T} \Smatrix^\dagger {\cal O} \Smatrix \DressedStateR{\psi}{T} 
= \langle {\psi}|e^{-i\SoftCharge{T}} \Smatrix^\dagger
{\cal O} 
\Smatrix e^{i\SoftCharge{T}}  |{\psi}\rangle \ .
\]
Notice that the state $\ket{\psi}$ now refers to a two-particle state (with each particle localised in appropriate wavepackets such that the state is in the domain of validity of the classical approximation).
Since this bare initial state is not an eigenstate of the soft charge while being an eigenstate of the hard charge, it is convenient to use the properties of the latter and rewrite this expectation value as
\[
O 
&=  \langle {\psi}|e^{iQ_h[T]} e^{-i(\SoftCharge{T}+Q_h[T])} \Smatrix^\dagger
{\cal O} 
\Smatrix e^{+i(\SoftCharge{T}+Q_h[T])}  e^{-iQ_h[T]} |{\psi}\rangle
\\
&=  \langle {\psi}|e^{iQ_h[T]} \Smatrix^\dagger
e^{-i(\SoftCharge{T}+Q_h[T])} 
{\cal O} 
e^{+i(\SoftCharge{T}+Q_h[T])} 
\Smatrix   e^{-i Q_h[T]} |{\psi}\rangle
\]
Thus, when evaluating final-state expectation values in a time-evolved dressed initial state, we may effectively take the initial state to be 
\[\label{eq:dressed2ptclState}
 e^{-i Q_h[T]}| \psi \rangle   &= \int \dd \Phi(p_1,p_2) \phi_b(p_1,p_2) e^{ -i Q_h[T]}| p_1 , p_2 \rangle 
% \\
% &\equiv \int \dd \Phi(p_1,p_2) \phi_b(p_1,p_2) e^{ -i \sum_{i=1}^2
% \HardCharge{f_{p_i}^{\helicity{}_i}}}| p_1 , p_2 \rangle
\]
while also conjugating the observable as
\[
{\cal O} \mapsto e^{-i(\SoftCharge{T}+Q_h[T])} 
{\cal O} 
e^{+i(\SoftCharge{T}+Q_h[T])}  \ .
\]
The wavepacket $\phi_b(p_1,p_2)$ appearing in Eq.~\eqref{eq:dressed2ptclState} localised both particles, and includes a translation factor $e^{ib\cdot p_1}$ which introduces the impact parameter. 
The contribution of the hard charge to classical observables can then be inferred by inspecting the action of the $S$-matrix on this effective state,
\[
\label{eq:PsiFin}
\Smatrix 
e^{- i Q_h[T]} | \psi \rangle \ ,
\]
which is aided by the fact that Fock states are eigenstates of the hard charge, see Eqs.~\eqref{eq:finalmassless} and \eqref{eq:Qhardmatter}-\eqref{eq:finalmassive}.
%
%%%%%%%%%%%%%%%%%%%%%%%%%
As discussed above, we further restrict our attention to the conservative sector, which allows us to obtain an explicit expression for this final state using eikonal methods, following the strategy of \cite{Cristofoli:2021jas}. 
%to calculate the action of the $S$-matrix on the  state $e^{- i Q_h} | \psi \rangle$. 
We will then see how the resulting final state can be used to re-sum the contributions of the hard charge to classical observables to all orders in the coupling.

Neglecting finite-energy radiation in the final state, we can express this state as 
\[
\Smatrix e^{-i Q_h[T]} | \psi \rangle = \int \dd \Phi(p_1,p_2,p_1',p_2') \phi_b(p_1,p_2)| p_1',p_2' \rangle \langle  p_1',p_2'| \, \Smatrix \,  e^{- i Q_h[T]} |  p_1 , p_2 \rangle \ .
\]
To proceed, we recall the action of the hard charge on the initial state derived in the previous section,
\[
Q_h  |  p_1 , p_2 \rangle  = \sum _{i\neq j}\left(Q_{h; i\ne j} (p_i, p_j, p'_j) + Q_{h; ii}(p_i, p'_i)\right)  |  p_1 , p_2 \rangle \equiv \sfQ_h[\{p\},\{p'\}] |  p_1 , p_2 \rangle \ ,
\]
where, for later convenience, we introduced the notation $\sfQ_h[\{p\},\{p'\}]$.
Note that, despite acting on the initial state, the hard charge retains information about the final state through the commutator \eqref{eq:Qhardmatter}. We can now isolate the non-trivial part of the $S$-matrix by subtracting the forward scattering contribution. This allows us to express the resulting state in terms of a $2\rightarrow 2$ scattering amplitude as 
\[
(\Smatrix-1) e^{-i Q_h[T]} | \psi \rangle = 
i  \int \dd \Phi(p_1,p_2,p_1',p_2') \phi_b(p_1,p_2)e^{-i \sfQ_h[\{p\},\{p'\}]
} \\
\times \mathcal{A}(p_1,p_2 \rightarrow p_1',p_2')\hdelta^4(p_1+p_2-p_1'-p_2') | p_1',p_2' \rangle \ . 
\]
This expression makes use of the exact $S$ matrix, and consequently the momentum mismatch $\sfq = p_1' - p_1 = p_2 - p_2'$ (we assume an elastic scattering process) \emph {needs not} be parametrically small. 
Using this mismatch as one of the integration variables allows us to eliminate the un-primed momentum variables in favour of the primed ones so that
%%%%%%%
\[\label{eq:FinalStateA4}
(\Smatrix-1) e^{-i Q_h}  | \psi \rangle  = i & \int \dd \Phi(p_1',p_2')  | p_1',p_2' \rangle 
\int \hd^4 \sfq \, \hdelta(2p'_1 \cdot \sfq - \sfq^2) \hdelta(2p_2'\cdot \sfq + \sfq^2) 
\\
&\times 
 \phi_b(p_1'-\sfq,p_2'+\sfq)
\exp\left( -i \sfQ_h[\{p'\}+\{\sfq\},\{p'\}]
%\sum_{i\neq j} \left(Q_{h; i\ne j} 
%(p'_i+\sfq_i, p'_j+\sfq_j, p'_j) 
%+ Q_{h; ii}(p'_i+\sfq_i, \sfq'_i)\right)
\right)\,\mathcal{A}_4(s,\sfq^2) \ , 
\]
where $\{p\}\equiv\{p_1, p_2\}=
\{p'_1+\sfq_1, p'_2+\sfq_2\}\equiv \{p'\}+\{\sfq\}$ and $\{\sfq\} = \{\sfq_1, \sfq_2\}$. 
In explicit expressions we will invoke momentum conservation and write $\sfq^\mu_1 = - \sfq^\mu$ and $\sfq_2^\mu = \sfq^\mu$.
We are now ready to express this result in terms of the eikonal phase $\chi(x_\perp;s)$ defined by the relation
\[\label{eq:EikonalDefinition}
i \hdelta \left(2 \bar{p}_1 \cdot \mathsf{q}\right) \hdelta\left(2 \bar{p}_2 \cdot \mathsf{q}\right) \mathcal{A}_4\left(s, \mathsf{q}^2\right)=
%\frac{1}{\hbar^4} 
\int \dd^4 x\,  e^{i \mathsf{q} \cdot x 
%/ \hbar
}\left\{e^{i \chi\left(x_{\perp} ; s\right) 
%/ \hbar
}-1\right\} \ .
\]
In this definition we are implicitly discarding quantum corrections, which modify this identity by introducing a remainder function $\Delta(x;s)$ multiplying the phase. 
We have also introduced $x_\perp$, which is a restriction of the Fourier variable $x^\mu$ to a two-dimensional plane defined by $x_\perp \cdot \bar{p}_1 = 0 = x_\perp \cdot \bar{p}_2$. 
Notice that this means that $x_\perp$ depends on $\sfq$.
An integration of the right-hand-side of Eq.~\eqref{eq:EikonalDefinition} over the two components of $x$ which are projected out of $x_\perp$ would recover the two delta functions on the left-hand-side.

Focusing on the classical eikonal, we apply the eikonal expression~\eqref{eq:EikonalDefinition} for the amplitude to Eq.~\eqref{eq:FinalStateA4} by defining the barred momentum variables to be
\[\label{eq:BarredMomenta}
\bar{p}_1 = p_1' -\frac{\sfq}{2} = p_1 +\frac{\sfq}{2} \ , 
\qquad
\bar{p}_2= p_2'+\frac{\sfq}{2} = p_2 - \frac{\sfq}{2} \ .
\]
These variables linearise the on-shell delta functions present in Eq.~\eqref{eq:FinalStateA4} so that we may directly take advantage of the eikonal resummation formula~\eqref{eq:EikonalDefinition}.
The final state becomes
\[
\Smatrix e^{-i Q_h} | \psi \rangle &= 
%\frac{1}{\hbar^4}  
\int \dd \Phi(p_1',p_2')  | p_1',p_2' \rangle \int \hd^4 \sfq\,  
\dd^4 x  \, \phi(p_1'-\sfq,p_2'+\sfq) e^{ip'_1 \cdot b_1} e^{i p_2' \cdot b_2 } e^{ - i \sfq \cdot (b_1 - b_2)
%/\hbar
}  e^{i \sfq \cdot x 
%/ \hbar
}\\
 &\times \exp\left(i \chi\left(x_{\perp} ; s\right)
 %/\hbar
 -  i \sfQ_h[\{p'\}+\{\sfq\},\{p'\}]
% \sum_{i\neq j} \left(Q_{h; i\ne j} 
% (p'_i+\sfq_i, p'_j+\sfq_j, p'_j) + 
% Q_{h; ii}(p'_i+\sfq_i, p'_i)\right)
 \right) \ ,
\]
where we have included the forward scattering contribution in this expression. We emphasise again that the barred momenta in Eq.~\eqref{eq:BarredMomenta} are entirely absorbed into the definition of $x_\perp$, which is restricted to lie in the plane orthogonal to $\bar{p}_1$ and $\bar{p}_2$, cf. Eq.~\eqref{eq:EikonalDefinition}. With this in mind, we continue to express the other quantities above in terms of the un-barred momenta $p_i'$ and $p_i$. 

We can now evaluate the integrals over $x$ and $\sfq$ using the saddle-point approximation, which localises the integrals on the coordinates $x_*$ and $\sfq_*$ defined as solutions of the saddle-point equations:
\begin{align}
\label{eq:SaddlePointEquations}
&\sfq^\mu =  - \frac{\partial}{\partial x_\mu} \chi\left(x_{\perp} ; s\right), \\
&x^\mu - b_{12}^\mu = -\frac{\partial}{\partial \sfq_\mu} \left(\chi\left(x_{\perp} ; s\right)-  
%\hbar 
\sfQ_h[\{p'\}+\{\sfq\},\{p'\}]
%\sum_{i\neq j} \left(Q_{h; i\ne j} 
%(p'_i+\sfq_i, p'_j+\sfq_j, p'_j) + 
%Q_{h; ii}(p'_i+\sfq_i, p'_i)\right)
\right) 
\ .
\nonumber
\end{align}
We simplify the first of these relations by noting that the eikonal phase $\chi$ for scalar field scattering is a function of $x_\perp^2$, which allows us to write
\[\label{eq:SaddlePointq}
\sfq_*^\mu = - 2 x_{\perp}^\mu \frac{\partial}{\partial x_\perp^2} \chi\left(x_{\perp} ; s\right) \ .
\]
To evaluate the second saddle point condition, we must resolve the dependence of $\chi\left(x_{\perp} ; s\right)$ on $\sfq$. This dependence is contained in the definition of $x_\perp^\mu$ through the conditions $x_\perp \cdot \bar{p}_i = 0$. We expose this dependence by decomposing $x^\mu$ into its orthogonal (to $\bar{p}_1$ and $\bar{p}_2$) and longitudinal components as
\[\label{eq:xdecomposition}
x^\mu  &= x_\perp^\mu + \alpha(\bar{p}_1 + \bar{p}_2)^\mu + \beta(\bar{p}_1 - \bar{p}_2)^\mu 
,\\
&
= x_\perp^\mu+\alpha(p_1'+p_2')^\mu + \beta(p_1'-p_2'-\sfq)^\mu \ .
\]
We have introduced expansion coefficients $\alpha$ and $\beta$ with generic $\sfq$-dependence. Differentiating with respect to $\sfq$ and noting that the left-hand side is independent of $q$, we obtain the following relation
\[\label{eq:Partialx_perpq}
\frac{\partial x_\perp^\mu}{\partial \sfq^\nu} = -\frac{\partial \alpha}{\partial \sfq^\nu}(\bar{p}_1 + \bar{p}_2)^\mu - \frac{\partial \beta}{\partial \sfq^\nu}(\bar{p}_1 - \bar{p}_2)^\mu + \beta\, \delta^\mu_\nu \ . 
\]
This determines the derivative of the eikonal,
\[
\label{eq:dChidq}
\frac{\partial}{\partial \sfq^\mu} \chi\left(x_{\perp} ; s\right) =  2 x_{\perp,\nu} \frac{\partial x_\perp^\nu}{\partial \sfq_{\mu}}\frac{\partial}{\partial x_\perp^2} \chi\left(x_{\perp} ; s\right) =  -\beta \, \sfq_{*,\mu} \ ,
\]
where we have used equations \eqref{eq:SaddlePointq} and \eqref{eq:Partialx_perpq} and $\sfq_* \cdot \bar{p}_i=0$ to arrive at the last equality. We can now express the second saddle point condition as
\[
x^\mu - b_{12}^\mu - \beta \sfq_*^\mu =    
%\hbar 
+\frac{\partial}{\partial \sfq_\mu}
\sfQ_h[\{p'\}+\{\sfq\},\{p'\}]
%\sum_{i\neq j} \left(Q_{h; i\ne j} 
%(p'_i+\sfq_i, p'_j+\sfq_j, p'_j) + 
%Q_{h; ii}(p'_i+ \sfq_i, p'_i)\right) 
\Big|_{\sfq=\sfq_*}
\ ,
\]
where we used Eq.~\eqref{eq:dChidq} to express the derivative of the eikonal in terms of $\sfq_*$, and the right-hand side is to be evaluated at $\sfq=\sfq_*$ after the derivative is taken.
The decomposition \eqref{eq:xdecomposition} allows us to rewrite the left-hand side as
\[
x^\mu - b_{12}^\mu - \beta \, \sfq^\mu &= x_\perp^\mu+\alpha(p_1'+p_2')^\mu + \beta(p_1'-p_2'-2\sfq)^\mu  - b_{12}^\mu, \\
&= x_\perp^\mu+\alpha(p_1+p_2)^\mu + \beta(p_1-p_2)^\mu  - b_{12}^\mu.
\]
We thus arrive at the relation
\[\label{eq:EikonalImpactParameter}
x_\perp^\mu - b_{12}^\mu &= - \alpha(p_1+p_2)^\mu - \beta(p_1-p_2)^\mu 
%\\
%&
+   
%\hbar 
\frac{\partial}{\partial \sfq_\mu}
\sfQ_h[\{p'\}+\{\sfq\},\{p'\}]
%\frac{\partial}{\partial \sfq_\mu}
%\sum_{i\neq j} \left(Q_{h; i\ne j} 
%(p'_i+\sfq_i, p'_j+\sfq_j, p'_j) + 
%Q_{h; ii}(p'_i+\sfq_i, p'_i)\right)
\Big|_{\sfq=\sfq_*} 
\ ,
\]
which describes the deviation of the eikonal impact parameter $x_\perp^\mu$ from $ b_{12}^\mu$ due to scattering. The coefficients $\alpha$ and $\beta$ can be deduced by contracting with $p_1$ and $p_2$ and noting that $p_i \cdot b_{12} = 0$. This gives 
\[\label{eq:alphabeta}
\alpha &= \frac{(m_2^2-m_1^2) \sfq_* \cdot x_\perp }{4 m_1^2m_2^2 - 4 (p_1 \cdot p_2){}^2} 
%-  
+
%\hbar 
\left(\frac{2(m_2^2- p_1 \cdot p_2)p_1^\mu}{4 m_1^2m_2^2 - 4 (p_1 \cdot p_2){}^2} + \frac{2(m_1^2- p_1 \cdot p_2)p_2^\mu}{4 m_1^2m_2^2 - 4 (p_1 \cdot p_2)^2}\right) \frac{\partial}{\partial \sfq^\mu} \mathsf{Q}_h\Big|_{\sfq=\sfq_*} \ , \\
\beta &= \frac{(p_1+p_2)^2 \sfq_* \cdot x_\perp }{4 m_1^2m_2^2 - 4 (p_1 \cdot p_2){}^2} 
%-  
+
%\hbar 
\left(\frac{2(m_2^2+ p_1 \cdot p_2)p_1^\mu}{4 m_1^2m_2^2 - 4 (p_1 \cdot p_2){}^2} - \frac{2(m_1^2+ p_1 \cdot p_2)p_2^\mu}{4 m_1^2m_2^2 - 4 (p_1 \cdot p_2) {}^2}\right) \frac{\partial}{\partial \sfq^\mu} \mathsf{Q}_h\Big|_{\sfq=\sfq_*}  \,
\]
%%%%%%%%%%%%%
where for brevity we used the shorthand notation $\mathsf{Q}_h\equiv \sfQ_h[\{p'\}+\{\sfq\},\{p'\}]$
for the action of the hard charge on a two-particle state with a specific parametrization of incoming momenta.

To compute the norm of the impact parameter and thus the right-hand side in Eq.~\eqref{eq:extraJloss}, we first derive relations between the scalar product of $x_\perp$ and other vectors,  
\[
&x_\perp \cdot b_{12} - b_{12}^2 =  
%\hbar 
\left(b_{12} \cdot \frac{\partial}{\partial \sfq}\right) \mathsf{Q}_h \Big|_{\sfq=\sfq_*} \ , \\
&x_\perp^2 - b_{12} \cdot x_\perp = \beta ( \sfq \cdot x_\perp) 
+
%\hbar
\left( x_\perp \cdot \frac{\partial}{\partial \sfq}\right) \mathsf{Q}_h
\Big|_{\sfq=\sfq_*} \ ,
\]
by contracting Eq.~\eqref{eq:EikonalImpactParameter} with $x_\perp$ and $b_{12}$. It is then not difficult to see that the squared impact parameter is 
\[
b_{12}^2 =& \, x_\perp^2 - \beta ( \sfq_* \cdot x_\perp)  
-
%\hbar 
\left( (b_{12} + x_\perp) \cdot \frac{\partial}{\partial \sfq} \right) \mathsf{Q}_h \Big|_{\sfq=\sfq_*} \\
 =& \, x_\perp^2 \left( 1- \frac{(p_1+p_2)^2 \sfq_*^2 }{4 m_1^2m_2^2 - 4 (p_1 \cdot p_2){}^2}\right)  
 -
 %\hbar 
 \left( (b_{12} + x_\perp) \cdot \frac{\partial}{\partial \sfq} \right) \mathsf{Q}_h \Big|_{\sfq=\sfq_*} \\
 &
 -
 %\hbar 
 ( \sfq_* \cdot x_\perp)\left(\frac{2(m_2^2+ p_1 \cdot p_2)p_1^\mu}{4 m_1^2m_2^2 - 4 (p_1 \cdot p_2){}^2} - \frac{2(m_1^2+ p_1 \cdot p_2)p_2^\mu}{4 m_1^2m_2^2 - 4 (p_1 \cdot p_2){}^2}\right) \frac{\partial}{\partial \sfq^\mu} \mathsf{Q}_h \Big|_{\sfq=\sfq_*} \ .
\]
In terms of the scattering angle $\Psi$, this reads 
\[
\label{eq:changeinbsq}
b_{12}^2 = & \, x_\perp^2 \cos^2\left( \frac{\Psi}{2}\right) 
-
%\hbar 
\left( (b_{12} + x_\perp) \cdot \frac{\partial}{\partial \sfq} \right) \mathsf{Q}_h \Big|_{\sfq=\sfq_*}\\
 & \, 
 -
 %\hbar 
 ( \sfq_* \cdot x_\perp)\left(\frac{2(m_2^2+ p_1 \cdot p_2)p_1^\mu}{4 m_1^2m_2^2 - 4 (p_1 \cdot p_2){}^2} - \frac{2(m_1^2+ p_1 \cdot p_2)p_2^\mu}{4 m_1^2m_2^2 - 4 (p_1 \cdot p_2){}^2}\right) \frac{\partial}{\partial \sfq^\mu} \mathsf{Q}_h \Big|_{\sfq=\sfq_*} \,.
\]
The first term on the right-hand side is the usual relation between the impact parameter $b_{12}$ measured at infinity and the eikonal impact parameter $x_\perp$, originally derived from a similar argument in Ref.~\cite{Cristofoli:2021jas}.
The other terms, proportional to the derivative of $\mathsf{Q}_h$, are the contribution of the coherent state dressing and are interpreted as the effect of changing the asymptotic frame as dictated by the coherent state shape.
Linearity on $\mathsf{Q}$ suggests that the linear composition of large gauge transformations is correctly captured by the hard charge we derived in the previous section.

While Eq.~\eqref{eq:changeinbsq} is exact, it is useful to expand it order by order in the generic coupling $g$ to compare with results existing in the literature for vector and graviton mediators. Doing so amounts to expanding only $Q_h$ and $q_*$ in that equation. 
Focusing on the vector precursor of that equation, Eq.~\eqref{eq:EikonalImpactParameter}, 
the leading order shift induced by the hard charge occurs at order $g^4$ and is equal to 
\[
\label{eq:g4shift}
x_\perp^\mu - b_{12}^\mu \Big|^{Q_h}_{g^4} = 
 \left(\frac{\partial}{\partial \sfq^\mu} \mathsf{Q}_h\right)\Big|_{(4)} \ .
\]
where the subscript denotes a restriction to the  $\mathcal{O}(g^4)$ contribution of the right-hand side and we have used the fact that $\partial \mathsf{Q}_h/\partial \sfq^\mu|_{\sfq=\sfq_*}$ is proportional to the tree-level impulse $\sfq_*^\mu$ at this order. 
This proportionality also implies that the combination
\[
\left(p_i \cdot\frac{\partial}{\partial \sfq} \right)\mathsf{Q}_h \Big|_{(4)} \sim g^2 \;p_i \cdot \sfq_*\Big|_{(2)} = 0 \ ,
\]
which thus drops out of the expression for $\alpha$ and $\beta$ in Eq.~\eqref{eq:alphabeta}.

Putting everything together, we find that the $\mathcal{O}(g^6)$ contribution to the impact parameter shift is
\[\label{eq:ImpactParameterNLO}
b_{12}^\mu-x_\perp^\mu\Big|^{Q_h}_{g^6} 
&=
 \frac{(m_2^2-m_1^2) (p_1+p_2)^\mu  }{4 m_1^2m_2^2 - 4 (p_1 \cdot p_2)^2} \;\sfq_*\Big|_{(2)} \cdot \left(\frac{\partial}{\partial \sfq}\mathsf{Q}_h\Big|_{\sfq=\sfq_*}\right)\Big|_{(4)}  \\
&+
(p_1+p_2)^\mu 
\left(\frac{2(m_2^2- p_1 \cdot p_2)p_1^\nu}{4 m_1^2m_2^2 - 4 (p_1 \cdot p_2){}^2} + \frac{2(m_1^2- p_1 \cdot p_2)p_2^\nu}{4 m_1^2m_2^2 - 4 (p_1 \cdot p_2)^2}\right) \left(\frac{\partial}{\partial \sfq^\nu} \mathsf{Q}_h\Big|_{\sfq=\sfq_*} \right)\Big|_{(6)} \\
&+ (p_1-p_2)^\mu \left(\frac{2(m_2^2+ p_1 \cdot p_2)p_1^\nu}{4 m_1^2m_2^2 - 4 (p_1 \cdot p_2){}^2} - \frac{2(m_1^2+ p_1 \cdot p_2)p_2^\nu}{4 m_1^2m_2^2 - 4 (p_1 \cdot p_2) {}^2}\right) \left(\frac{\partial}{\partial \sfq^\nu} \mathsf{Q}_h\Big|_{\sfq=\sfq_*} \right)\Big|_{(6)}\\
&+  \frac{(p_1+p_2)^2 (p_1-p_2)^\mu  }{4 m_1^2m_2^2 - 4 (p_1 \cdot p_2){}^2} \;\sfq_*\Big|_{(2)} \cdot \left(\frac{\partial}{\partial \sfq}\mathsf{Q}_h\Big|_{\sfq=\sfq_*}\right)\Big|_{(4)} - \left(\frac{\partial}{\partial \sfq^\mu} \mathsf{Q}_h\Big|_{\sfq=\sfq_*} \right)\Big|_{(6)} ,
\]
where we have implicitly used the order $g^4$ result for $x_\perp$ given in Eq.~\eqref{eq:g4shift}. We stress that the complete impact parameter shift also includes the standard eikonal part, carried by the parameters $\alpha$ and $\beta$ in Eq.~\eqref{eq:alphabeta} or the first term on the right-hand side of Eq.~\eqref{eq:changeinbsq}, in addition to the $Q_h$-dependent parts given through ${\cal O}(g^6)$ by Eqs.~\eqref{eq:g4shift} and \eqref{eq:ImpactParameterNLO}. 

\section{Asymptotic charges 
for the canonical $\longleftrightarrow$ intrinsic relation
}
\label{sec:VV}

\begin{figure}
    \centering
    \includegraphics[width=0.9\linewidth]{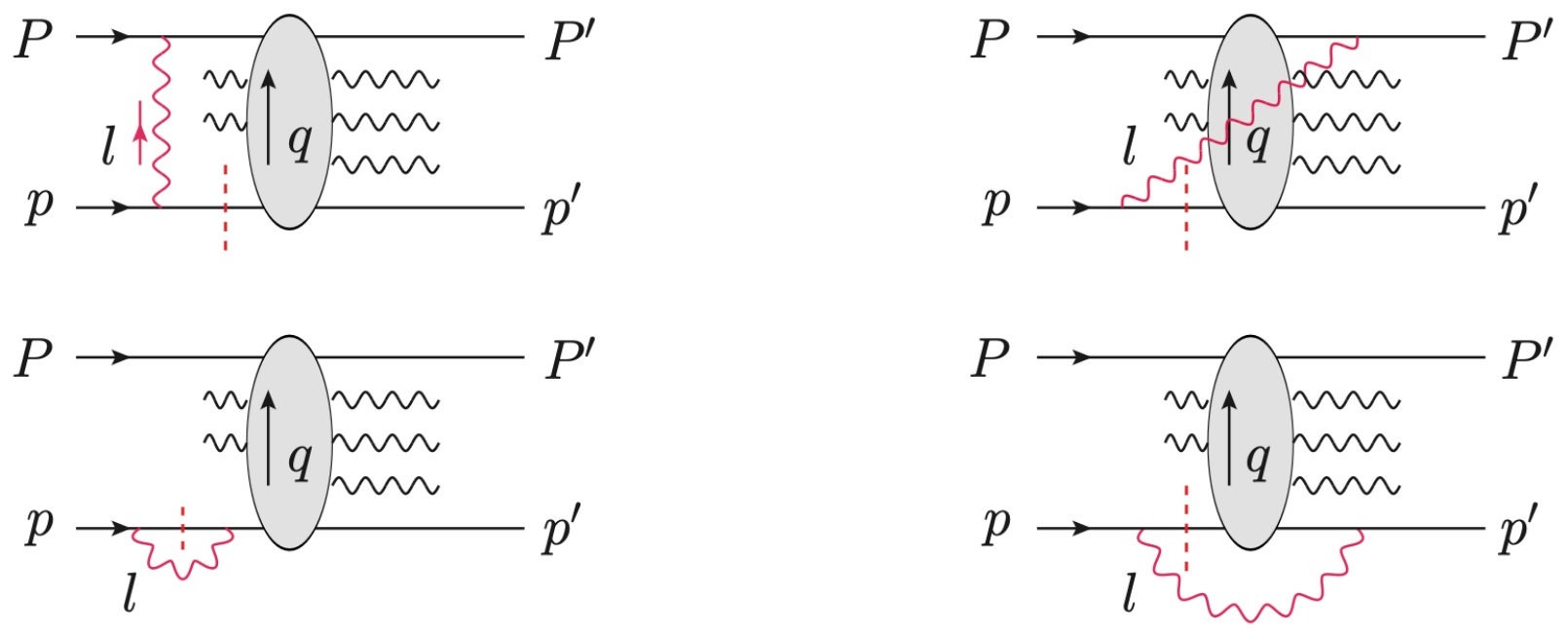}
    \caption{The diagrams contributing to the commutator $[\Smatrix, Q_s]$ for a process with 4 matter lines. The line crossed by the dashed line is cut; 
     the red wiggly line denotes a cut mediator propagator. The cut propagator originates from the delta function in the definition of the soft charge, Eq.~\eqref{eq:defOfSoftCharge}. It is a particular realization of the cut red line in Fig.~\ref{fig:4matter_general}.}
    \label{fig:4matter}
\end{figure}

In section~\ref{sec:general_hard_charges} we discussed general supertranslations from an on-shell perspective, and in section~\ref{sec:consequences} we understood their effect on the impact parameter at infinity in a conservative scattering process.
We now specialize to the particular supertranslation
which corresponds to the shape function defined in the notation of Eq.~\eqref{eq:coherent_dressing} by
\[
\label{eq:VVdressing}
F^\helicity_p(k) = \frac{1}{2}\hdelta(2 p \cdot k)  %\mathcal{A}_3(p+k \rightarrow p, k_\helicity) 
\SoftOT{p+k}{p}{k_\helicity} \ .
\]
We will obtain explicit results for the correction to the impact parameter in a (conservative, two-particle) scattering process dressed with this choice of $F^\helicity_p$, and comment on consequences for the angular momentum loss and related observables.

The supertranslation associated to Eq.~\eqref{eq:VVdressing} is particularly important as it connects~\cite{Veneziano:2022zwh,Elkhidir:2024izo,Menezes:2026edi,Menezes:2026fmj} the canonical and intrinsic frames which are prominent in different approaches to the gravitational two-body problem. Note that the factor $\SoftOT{a}{b}{c}$ appearing in Eq.~\eqref{eq:VVdressing} is the $(a\rightarrow b,c)$ soft factor. 
There is one such factor for each incoming matter particle of momentum $p$.
Between the phase space integral in the definition of the soft charge and the delta function in Eq.~\eqref{eq:VVdressing}, 
\begin{equation}
\label{eq:defOfSoftCharge}
Q_s \equiv \frac{1}{2}\sum_\helicity \int \dd\Phi(k) \hdelta(2 p \cdot k)  %\mathcal{A}_3(p+k \rightarrow p, k_\helicity) 
\SoftOT{p+k}{p}{k_\helicity}
a^\dagger_\helicity(k)  
+ \textrm{h.c.} \ ,
\end{equation}
the integrand is effectively proportional to $\delta(\omega)$, so the coherent state dressing is of the general type discussed earlier in~Sec.~\ref{sec:general_hard_charges}.

We may proceed by either directly applying the general expressions derived in Sec.~\ref{sec:general_hard_charges}, or by simply re-deriving the hard charge 
independently, making use of the manifest Lorentz-invariance of Eq.~\eqref{eq:defOfSoftCharge}.
We will do the latter in this section, and then use the results to discuss the consequences of this BMS supertranslation for the impact parameter. 
As in Sec.~\ref{sec:general_hard_charges} and in contrast with Ref.~\cite{Elkhidir:2024izo} here we keep the exact dependence on the momentum transfer, which is finite unlike its perturbative counterpart which is ${\cal O}(\hbar)$. This is important because, as we discussed, in the definition of the asymptotic states the momentum transfer is effectively an integration variable; the ensuing (eikonal) resummation relates it to the impulse.

As we saw in the general discussion of Sec.~\ref{sec:general_hard_charges}, the modes $\ell=0, 1$ are naturally projected out from the supertranslation parameter. Shifts of the supertranslation parameter by such modes are local symmetries of supertranslations, i.e. they are coordinate transformations that leave the metric coefficients invariant. 
For the purpose of the Veneziano-Vilkovisky supertranslation analysed in this section, it is convenient to forego carrying out this projection explicitly.

\subsection{Exact hard charges for the canonical $\longleftrightarrow$ intrinsic frame connection}
\label{sec:VVhardcharges}

\noindent{}
{\it Massless hard charge:} 
The $C_0$ contribution to the commutator $[\Smatrix, Q_s]$ was discussed at length in \cite{Elkhidir:2024izo}, and we will only sketch it here. The contribution to the commutator $[\Smatrix, Q_s]$ from the soft graviton attached to an incoming graviton is
\begin{align}
\label{eq:CmasslessFull}
C_0 =& \frac{1}{2}
\int \dd\Phi(l,P',P)\,\hdelta(2p\cdot l)  \, (i H(P,P'))  \left[\; - \frac{X_0}{2K\cdot l + i \varepsilon} \right]
\\
=&  \frac{1}{2}\int \dd\Phi(P',P)\, (i H(P,P'))  \left[\; - X_0 I(p, K) \, \right] \ ,
\end{align}
where the factor $X_0$, with the subscript indicating the masslessness of the external mediator momentum $K$, is  
\begin{align}
X_0 &= \sum_\helicity  
%\mathcal{A}_3(P, k_\helicity  \rightarrow P+k)  
\SoftTO{K}{l_\helicity}{K+l}  
%\mathcal{A}_3(p + k \rightarrow p, k_\helicity )  
\SoftOT{p + l}{p}{l_\helicity}  \ .
%%\\
%%X_0'&= \sum_\helicity  
%%%\mathcal{A}_3(P'-k, k_\helicity  \rightarrow P') 
%%\SoftTO{K-l}{k_\helicity}{K}  
%%%\mathcal{A}_3(p +k \rightarrow p, k_\helicity)
%%\SoftOT{p + l}{p}{l_\helicity} 
%%%=  \kappa^2 \left((P'\cdot p)^2 -\frac{1}{2} P'{}^2 p^2)\right)
%%\, ,
\end{align}
Since it depends only on one matter momentum --- the one on which the soft charge is attached to --- its evaluation does not refer to the momentum transfer and thus the result of Ref.~\cite{Elkhidir:2024izo},
%\[
%Q_{h, 0} = -2G \, \sum_{i=1}^4 \sum_\helicity  \int \dd \Phi(k) \, 
%p \cdot k \Big[
%\log  \left(\frac{p\cdot k}{m k^0} \right)+Y(\beta, \epsilon)\Big] \, 
%a^\dagger_\helicity (k) a_\helicity (k) \, .
%\]
\[
Q_{h, 0} &= \sum_\helicity \int \dd \Phi(k) \, Q^0_{h}(k) \, a^\dagger_\helicity (k) a_\helicity (k) \,,
\\
Q^0_{h}(k) &=  -2G \, \sum_{i}  \, 
%{\cal P}^{\ell\ge 2}\, 
p_i \cdot k \Big[
\log  \left(\frac{p_i\cdot k}{m_i k^0} \right)+Y(\beta, \epsilon)\Big] \,,
\label{eq:VVdirect}
\]
is exact.\footnote{Note that the action of the number operator on an operator with momentum $K$ will evaluate its coefficient on the momentum $K$.} The relevant integral can be either evaluated directly, or obtained as the massless limit of $I(p, P)$ in Eq.~\eqref{eq:OffDiagIntegral} below. 
We assumed here that all external mediators are identical and the index $i$ runs over all the external matter particles. 

It is not difficult to see that the sum over matter particles in Eq.~\eqref{eq:VVdirect} eliminates the $Y(\beta, \epsilon)$ constant because $\sum_i p_i\cdot k = -k^2 = 0$. 
The remainder of that equation reproduces the supertranslation parameter of Ref.~\cite{Veneziano:2022zwh}, and they both differ 
from Eqs.~\eqref{eq:IK}-\eqref{eq:finalmassless} by $\ell=0, 1$ modes, i.e. by a gauge symmetry of supertranslations, as expected. Enforcing that projection eliminates terms from the formal expansion of the logarithm in $\bm u_i\cdot \bm n$; for the rest of this section it is convenient to choose the ``supertranslation gauge" so that these terms are kept in the parameter.

\smallskip

\noindent{}
{\it Massive hard charge:} 
The matter contribution, $C_M$, to the commutator is more interesting.
Our assumptions imply that the massive particle is present in both the initial and final state, and therefore there are two related contributions of interest, shown on the same row in Fig.~\ref{fig:4matter} for an $S$-matrix element with four external matter particles. Compared to Fig.~\ref{fig:4matter_general} both ends of the soft graviton have the same type of vertex --- a soft factor. 
As in the general case, it is convenient to discuss them one pair at a time.

The two contributions to $C_M$ contain the sums 
\begin{align}
\label{eq:XandXprime}
X &= \sum_\helicity  
%\mathcal{A}_3(P, k_\helicity  \rightarrow P+k)  
\SoftTO{P}{l_\helicity}{P+l}  
%\mathcal{A}_3(p + k \rightarrow p, k_\helicity )  
\SoftOT{p + l}{p}{l_\helicity}  
%= \kappa^2 \left((P\cdot p)^2 -\frac{1}{2} P{}^2 p^2)\right)
\, ,
\\
X'&= \sum_\helicity  
%\mathcal{A}_3(P'-k, k_\helicity  \rightarrow P') 
\SoftTO{P'-l}{l_\helicity}{P'}  
%\mathcal{A}_3(p +k \rightarrow p, k_\helicity)
\SoftOT{p + l}{p}{l_\helicity} 
%=  \kappa^2 \left((P'\cdot p)^2 -\frac{1}{2} P'{}^2 p^2)\right)
\, ,
\end{align}
written here with the complete momentum dependence.

We will be interested in the exact state after the time evolution.
Since in that state the change in the momentum is not ${\cal O}(\hbar)$, we keep the exact dependence; we avoid taking explicit references to the momentum exchanged with other particles by phrasing the calculation only in terms of momenta external to the process. 
If the endpoints of the zero-frequency graviton are attached to different particles, the commutator contribution is 
\begin{align}
\label{eq:CmassiveFull}
C_M =& \frac{1}{2}
\int \dd\Phi(l,P',P)\,\hdelta(2p\cdot l)  \, (i H(P,P'))  \left[\; - \frac{X'}{-2P'\cdot l + i \varepsilon} - \frac{X}{2P\cdot l + i \varepsilon} \right]
\\
=& \frac{1}{2} \int \dd\Phi(P',P)\, (i H(P,P'))  \left[\; X' I(p, P') - X I(p, P) \, \right] \ .
\end{align}
If they are attached to the same particle, the commutator contribution is
\begin{align}
\label{eq:Cmassive1D}
C_M^{\text{diag.}} =& \frac{1}{2} \int \dd\Phi(l,P',P)\,\hdelta(2p\cdot l)  \, (i H(P,P'))  \left[\; - \frac{X'_\text{diag}}{-2p'\cdot l + i \varepsilon} 
-\frac{X_\text{diag}}{p^2 -m^2 + i \varepsilon}\right]
\\
=& \frac{1}{2} \int \dd\Phi(P',P)\, (i H(P,P'))  \left[\;  {X'_\text{diag}}\, I(p, p') - 
%\frac{X_\text{diag}}{p^2 -m^2 + i \varepsilon} 
X_\text{diag} J_0(p)\right] \ ,
\end{align}
where we defined 
\begin{align}
\label{eq:Defintegral}
I(p, P) &= \int \dd\Phi(l)\,\frac{ \hdelta(2p\cdot l)  }{2 P\cdot l \pm i \varepsilon}   \,,
\\
J_0(p) & \equiv \lim_{\beta\rightarrow 0} 
\frac{1}{p^2-m^2} \int \dd\Phi(l)\, \hdelta((p-l)^2-m^2) 
%\delta((p-l)^2-m^2) 
\Big|_{p\rightarrow p+\beta Y} \ .
\end{align}
While $X$ and $X'$ appear to depend on the integration variable $l$, its softness 
$l\rightarrow 0$ implied by the phase space condition and the delta function evaluates them at $l=0$ and thus allows us to pull them out of the integrals.

Thus, following the same pattern as in Sec.~\ref{sec:general_hard_charges}, the matter contribution $C_M$ to the commutator $[\Smatrix, Q_s]$ can be written as
\begin{equation}
\label{eq:QhardmatterVV}
C_M = -[\Smatrix, Q_{h, M}] \,,
\qquad Q_{h, M} = \sum_i \int \dd\Phi(p_i)\,  Q^M_{h, i} \, A^\dagger_i(p_i) A_i(p_i) \ ,
\end{equation}
where the sum is over all the types of matter particles, see footnote~\ref{foot:footnote9}.
$Q^M_{h, i}$ is a sum over all the incoming matter particles (because $k$ can attach to all matter particles and in $C_M$ we looked at an incoming particle and its outgoing counterpart). It is
\begin{align}
Q^M_{h, i} =  \frac{1}{2} \bigg(\sum_{j\ne i} & \left[X'(p_i, p_j') I(p_i, p'_j) - X(p_i, p_j) I(p_i, p_j) \right]
\cr
&
+ X'_\text{diag}(p_i, p_i') I(p_i, p'_i) - X_\text{diag}(p_i, p_i) J_0(p_i) \bigg)\,.
\end{align}
Splitting this into two-particle contributions, $Q^M_{h;ij}$ with $i\ne j$, and  single-particle ones, $Q_{h;ii}$, we have for these quantities:
\begin{align}
Q^M_{h; i\ne j}(p_i, p_j, p'_j) &=  \frac{1}{2} \left( X'(p_i, p_j') I(p_i, p'_j) - X(p_i, p_j) I(p_i, p_j) \right) \,,
\\
Q^M_{h; ii}(p_i, p'_i) &=  \frac{1}{2} \left( X'_\text{diag}(p_i, p_i') I(p_i, p'_i) - X_\text{diag}(p_i, p_i) J_0(p_i) \right) \,.
\end{align}

Using the $\beta$ regularization, i.e. $p\rightarrow p + m \beta P/P^0$ in the quadratic form $(p-k)^2 = m^2$ of the cut condition, dropping terms proportional to $\beta^2$ and sending $\beta\rightarrow 0$ at the end, it is not difficult to find that ($Z=\sqrt{(p\cdot P)^2-p^2P^2}$)
\begin{align}
\label{eq:OffDiagIntegral}
I(p, P) &= 
%\frac{1}{(2\pi)^2}\frac{\pi}{4Z}\ln \frac{p\cdot P + Z }{p\cdot P - Z} =
\frac{1}{16\pi Z}
\ln \frac{p\cdot P + Z }{p\cdot P - Z} \,,
\\
\label{eq:DiagIntegral}
J_0(p) & \equiv \lim_{\beta\rightarrow 0} \frac{1}{p^2-m^2} \int \dd\Phi(l) \, 
\hdelta((p-l)^2-m^2) 
%\delta((p-l)^2-m^2) 
\Big|_{p\rightarrow p+\beta Y} = 
\frac{1}{8
\pi p^2} \ .
\end{align}
In the $\beta$ regularization the second integral is ${\cal O}(\beta)$, but the dangling propagator is $1/(p^2-m^2)\sim 1/\beta$, so the contribution is finite.

We note here that, in the limit $P^2\to 0$, the integral $I(p, P)$ becomes the integral determining the massless hard charge in an off-shell regularization; it is not difficult to see that the relation
\[
Y(\epsilon, \beta)\mapsto \ln P^2 
\]
is the map between this off-shell regulator and the $\beta$ regulator of Refs.~\cite{Elkhidir:2024izo, Bini:2024rsy}. This divergence cancels in the massless hard charge upon summation over the external matter particles, as discussed below Eq.~\eqref{eq:VVdirect}.

\subsection{Hard charges for the canonical $\longleftrightarrow$ intrinsic frame connection for various mediators}

In this section we illustrate the consequences of supertranslations on observables in theories with mediators of various spins by studying the 
particular transformation connecting the canonical and intrinsic frames.
To this end we will need the action of $Q_{h, M}$ on two-particle states.
We collect these expressions here, for scalar, abelian vector and graviton mediators, and utilize them in Sec.~\ref{sec:changeINb} to derive the change in impact parameter.

We stress that the $\sfq$-expansion below is an expansion of the specific canonical $\leftrightarrow$ intrinsic dressing on a two-particle state, not of the hard-charge construction itself. The all-orders relation \eqref{eq:changeinbsq} remains exact; Eqs.~\eqref{eq:LOHardChargeScalar}, \eqref{eq:LOHardChargeQED}, and \eqref{eq:LOHardChargeGR}
provide its leading PM truncation, while the closed-form integrals $I(p,P)$ and $J_0(p)$ in Eqs.~\eqref{eq:OffDiagIntegral}-\eqref{eq:DiagIntegral} make the all-orders 
expression available in principle.

\subsubsection{Scalar theory, $2\rightarrow 2$ scattering}
\label{sec:scalars}

Our discussion of the hard charges relies solely on the  
three-point amplitudes of the theory. The details of higher-point vertices and their contribution to the classical limit are captured by the exact impulse. 
Asymptotic symmetries for a theory with scalar mediators and only trilinear couplings were discussed in Ref.~\cite{Campiglia:2018see}. Assuming such a theory, 
i.e. assuming only a trilinear interaction between massive scalars and the massless mediator scalar, fixes the $X$ and $X'$ factors in Sec.~\ref{sec:VVhardcharges} to be constants: 
\begin{align}
X&= \sum_\helicity  
%\mathcal{A}_3(P, k_\helicity  \rightarrow P+k)  
\SoftTO{P}{k_\helicity}{P+k}  
%\mathcal{A}_3(p + k \rightarrow p, k_\helicity )  
\SoftOT{p + k}{p}{k_\helicity}  = e_1 m_1 \, e_2 m_2  \ ,  
\\
X'&= \sum_\helicity  
%\mathcal{A}_3(P'-k, k_\helicity  \rightarrow P') 
\SoftTO{P'-k}{k_\helicity}{P'}  
%\mathcal{A}_3(p +k \rightarrow p, k_\helicity)
\SoftOT{p + k}{p}{k_\helicity}= e_1 m_1 \, e_2 m_2 \ , 
\\
X_\text{diag}&= e_1^2 m_1^2 \,,
\qquad
X_\text{diag}'= e_1^2 m_1^2 \,,
\end{align}
and similar expressions but with $e_1\rightarrow e_2$ and $m_1\rightarrow m_2$ when $X_\text{diag}$ and $X_\text{diag}'$ correspond to interactions of particle $2$.
Here, the quantities $e_1$ and $e_2$ are dimensionless coupling constants, having scaled the appropriate masses out from the relevant three-point amplitudes. 
Exposing the momentum transfer/integration variable/impulse in the eikonal asymptotic states, as in Sec.~\ref{sec:consequences}, puts the hard charge in the form
\begin{align}
Q^\text{sc}_h|1,2\rangle 
%&=\left( Q^\text{sc}_{h, 1 2}(p_1, p_2, p'_2) 
%+ Q^\text{sc}_{h, 2 1}(p_2, p_1, p'_1) 
%+ Q^\text{sc}_{h, 1 1}(p_1, p'_1) 
%+ Q^\text{sc}_{h, 2 2}(p_2, p'_2) \right)  |1,2\rangle
%\\
&= \left(Q^\text{sc}_{h, 1 2}(p'_1-\sfq, p'_2+\sfq, p'_2) + Q^\text{sc}_{h, 2 1}(p'_2+\sfq, p'_1-\sfq, p'_1) \right.
\cr
&~~\left. + Q^\text{sc}_{h, 1 1}(p'_1-\sfq, p'_1) + Q^\text{sc}_{h, 2 2}(p'_2+\sfq, p'_2)   \right)  |1,2\rangle \,.
\end{align}
Using the results of Sec.~\ref{sec:VVhardcharges} and formally expanding\footnote{\label{foot:qexpansion} Since $\sfq$ is the impulse, this expansion is equivalent to the expansion in the coupling, i.e. it is the standard PM expansion, and should not be confused with the expansion in the soft region corresponding to the classical limit.} in $\sfq$ shows that the off-diagonal and diagonal parts are
\begin{align}
&Q^\text{sc}_{h, 1 2}(p'_1-\sfq, p'_2+\sfq, p'_2) + Q^\text{sc}_{h, 2 1}(p'_2+\sfq, p'_1-\sfq, p'_1) \cr
&\qquad\qquad\qquad\qquad
= -\frac{1}{16\pi}\frac{e_1e_2}{m_1 m_2}\left[-\frac{1}{\gamma^2 v^2} + \frac{ {\rm arctanh}(v)}{\gamma^2 v^3}\right] \sfq^2 + {\cal O}(\sfq^4) \,,
\\
&Q^\text{sc}_{h, 1 1}(p'_1-\sfq, p'_1) + Q^\text{sc}_{h, 2 2}(p'_2+\sfq, p'_2) \cr
&\qquad\qquad\qquad\qquad
= \frac{m_1 m_2}{16\pi}\left[
\frac{1}{6} \left(\frac{e_1^2}{ m_1^4} + \frac{e_2^2}{ m_2^4}  \right) \sfq^2 \right] + {\cal O}(\sfq^4) \ ,
\end{align}
where $\gamma=p_1\cdot p_2/(m_1 m_2)$.
Thus, the action of the total massive hard charge for scalar mediators on a bare two-particle state is, to leading nontrivial order in the small transferred momentum expansion, 
\begin{align}\label{eq:LOHardChargeScalar}
Q^\text{sc}_h|1,2\rangle &=  \frac{e_1 e_2}{16\pi m_1 m_2} 
\left[ 
\left(\frac{1}{6} \left(\frac{e_1 m^2_2}{ e_2 m^2_1} + \frac{e_2 m^2_1}{ e_1 m^2_2}  \right) +\frac{1}{\gamma^2 v^2} - \frac{ {\rm arctanh}(v)}{\gamma^2 v^3}\right) \sfq^2+{\cal O}(\sfq^4)\right] |1,2\rangle\ .
\end{align}
%
%Only the second term contributes to the saddle-point approximation for the evolution of $e^{-i Q_h} |\psi\rangle$; the constant term, proportional to $(e_1^2+e_2^2)$, goes along for the ride and cancels in expectation values since it is only a constant phase.

\subsubsection{QED, $2\rightarrow 2$ scattering}

Asymptotic symmetries for theories with vector mediators were discussed in Ref.~\cite{Campiglia:2015qka, Campiglia:2018dyi}. 
For scalar QED the $X$ and $X'$ factors in Sec.~\ref{sec:VVhardcharges} are:  
\begin{align}
X&= \sum_\helicity  
%\mathcal{A}_3(P, k_\helicity  \rightarrow P+k)  
\SoftTO{P}{k_\helicity}{P+k}  
%\mathcal{A}_3(p + k \rightarrow p, k_\helicity )  
\SoftOT{p + k}{p}{k_\helicity}  = -4 e_1 e_2 \, P \cdot p  \ ,  
\\
X'&= \sum_\helicity  
%\mathcal{A}_3(P'-k, k_\helicity  \rightarrow P') 
\SoftTO{P'-k}{k_\helicity}{P'}  
%\mathcal{A}_3(p +k \rightarrow p, k_\helicity)
\SoftOT{p + k}{p}{k_\helicity}= -4 e_1 e_2 \, P' \cdot p \ , 
\\
X_\text{diag}&= -4 e_1^2 \,  p^2  \,,
\qquad
X_\text{diag}'= -4 e^2_1  \, p' \cdot p \, ,
\end{align}
and similar expressions with $e_1\rightarrow e_2$ when $X_\text{diag}$ and $X_\text{diag}'$ correspond to interactions of particle $2$.
Exposing, as in the scalar case, the dependence on the momentum transfer/integration variable/impulse  puts the matter hard charge in the form
\begin{align}
Q^\text{qed}_h|1,2\rangle 
%&=\left( Q^\text{qed}_{h, 1 2}(p_1, p_2, p'_2) 
%+ Q^\text{qed}_{h, 2 1}(p_2, p_1, p'_1) 
%+ Q^\text{qed}_{h, 1 1}(p_1, p'_1) 
%+ Q^\text{qed}_{h, 2 2}(p_2, p'_2) \right)  |1,2\rangle
%\\
&= \left(Q^\text{qed}_{h, 1 2}(p'_1-\sfq, p'_2+\sfq, p'_2) + Q^\text{qed}_{h, 2 1}(p'_2+\sfq, p'_1-\sfq, p'_1) \right.
\cr
&~~~\left. + Q^\text{qed}_{h, 1 1}(p'_1-\sfq, p'_1) + Q^\text{qed}_{h, 2 2}(p'_2+\sfq, p'_2)   \right)  |1,2\rangle \,.
\end{align}
Using the results of Sec.~\ref{sec:VVhardcharges} and formally expanding in $\sfq$ \footnote{See footnote~\ref{foot:qexpansion} for the interpretation of this expansion.} gives the off-diagonal and diagonal parts as
\begin{align}
Q^\text{qed}_{h, 1 2}(p'_1-\sfq, p'_2+\sfq, p'_2) &+ Q^\text{qed}_{h, 2 1}(p'_2+\sfq, p'_1-\sfq, p'_1) \cr
&=\frac{1}{4\pi}\frac{e_1e_2}{m_1 m_2}\left[-\frac{1}{\gamma v^2} + \frac{ {\rm arctanh}(v)}{\gamma^3 v^3}\right] \sfq^2 + {\cal O}(\sfq^4)
\\
Q^\text{qed}_{h, 1 1}(p'_1-\sfq, p'_1) + &Q^\text{qed}_{h, 2 2}(p'_2+\sfq, p'_2)  
= \frac{1}{4\pi}\left[
%-2 
\frac{1}{3} \left(\frac{e_1^2}{ m_1^2} + \frac{e_2^2}{ m_2^2}  \right) \sfq^2 \right] + {\cal O}(\sfq^4) \, .
\nonumber
\end{align}
Collecting the pieces yields the action of the scalar-QED hard matter
charge on a two-particle state without the coherent state dressing. 
At leading nontrivial order in the post-Minkowskian expansion, it is 
\begin{align}\label{eq:LOHardChargeQED}
Q^\text{qed}_h|1,2\rangle &=  \frac{e_1 e_2}{4\pi m_1 m_2} 
\left[ 
\left(\frac{1}{3} \left(\frac{e_1 m_2}{ e_2 m_1} + \frac{e_2 m_1}{ e_1 m_2}  \right) -\frac{1}{\gamma v^2} + \frac{ {\rm arctanh}(v)}{\gamma^3 v^3}\right) \sfq^2+{\cal O}(\sfq^4)\right] |1,2\rangle\ .
\end{align}
%
%As in the case of scalar mediators, only the second term contributes to the saddle-point approximation for the evolution of $e^{-i Q_h} |\psi_{\rm in}\rangle$; the constant term, proportional to $(e_1^2+e_2^2)$, goes along for the ride and cancels in expectation values since it is only a constant phase.

\subsubsection{GR, $2\rightarrow 2$ scattering}

Finally, we reach the case of graviton mediators. The factors $X$ and $X'$ in Sec.~\ref{sec:VVhardcharges} are 
\begin{align}
\label{eq:XandXprimeGR}
X &= \sum_\helicity  
%\mathcal{A}_3(P, k_\helicity  \rightarrow P+k)  
\SoftTO{P}{k_\helicity}{P+k}  
%\mathcal{A}_3(p + k \rightarrow p, k_\helicity )  
\SoftOT{p + k}{p}{k_\helicity}  = \kappa^2 \left((P\cdot p)^2 -\frac{1}{2} P{}^2 p^2 \right)
\, ,
\\
X'&= \sum_\helicity  
%\mathcal{A}_3(P'-k, k_\helicity  \rightarrow P') 
\SoftTO{P'-k}{k_\helicity}{P'}  
%\mathcal{A}_3(p +k \rightarrow p, k_\helicity)
\SoftOT{p + k}{p}{k_\helicity} =  \kappa^2 \left((P'\cdot p)^2 -\frac{1}{2} P'{}^2 p^2 \right)
\, ,
\\
X_\text{diag}&=X\Big|_{P=-p}=\kappa^2\; \frac{p^4}{2} \,, 
\qquad
X'_\text{diag}=X'\Big|_{P'=p'}= \kappa^2 \left((p'\cdot p)^2 -\frac{1}{2} p'{}^2 p^2 \right) \,.
\end{align}
We organize the matter hard charge as in the case of scalar and vector mediators, 
\begin{align}
Q_h|1,2\rangle 
%&=\left( Q_{h, 1 2}(p_1, p_2, p'_2) 
%+ Q_{h, 2 1}(p_2, p_1, p'_1) 
%+ Q_{h, 1 1}(p_1, p'_1) 
%+ Q_{h, 2 2}(p_2, p'_2) \right)  |1,2\rangle
%\\
&= \left(Q_{h, 1 2}(p'_1-\sfq, p'_2+\sfq, p'_2) + Q_{h, 2 1}(p'_2+\sfq, p'_1-\sfq, p'_1) \right.
\cr
&~~~\left. + Q_{h, 1 1}(p'_1-\sfq, p'_1) + Q_{h, 2 2}(p'_2+\sfq, p'_2)   \right)  |1,2\rangle \ .
\end{align}
Then, the formal expansion in $\sfq$ \footnote{See footnote~\ref{foot:qexpansion} for the interpretation of this expansion.}  gives the off-diagonal and diagonal parts of the hard charge to leading PM order \footnote{That is ${\cal O}(G^3)$, with one factor of Newton's constant from the explicit $\kappa^2$ and two more from the two factors of the impulse $\sfq = \Delta p$.}
\begin{align}
Q_{h, 1 2}(p'_1-\sfq, p'_2+\sfq,& p'_2) + Q_{h, 2 1}(p'_2+\sfq, p'_1-\sfq, p'_1) \\
&= -\frac{\kappa^2}{64\pi} \sfq^2 \left[-2-\frac{2}{v^2} + \frac{2(1-3 v^2){\rm arctanh}(v)}{v^3} \right]+{\cal O}(\sfq^4) \,,
\nonumber
\\
Q_{h, 1 1}(p'_1-\sfq, p'_1) &+ Q_{h, 2 2}(p'_2+\sfq, p'_2) 
%\\&\!\!\!\!\!\!\!\!
= -\frac{\kappa^2}{64\pi} \left[ \frac{22}{3} \sfq^2\right] 
 +{\cal O}(\sfq^4) \,,
\end{align}
leading to 
\begin{align}\label{eq:LOHardChargeGR}
Q_h|1,2\rangle = - \frac{\kappa^2}{64\pi} \left[ 
\sfq^2\left(\frac{16}{3}-\frac{2}{v^2} + \frac{2(1-3 v^2){\rm arctanh}(v)}{v^3} \right)+{\cal O}(\sfq^4)\right] |1,2\rangle
\end{align}
for the action of the total matter hard charge on a bare 2-particle state, to leading nontrivial order in the expansion in the small momentum transfer, as for spin-0 and spin-1 mediators.
%
%Only the second term contributes to the saddle-point approximation for the evolution of $e^{-i Q_h} |\psi_{\rm in}\rangle$; the constant term, $\kappa^2 m_i^2$ goes along for the ride and cancels in expectation values since it is only a phase.
        
\subsection{Change in impact parameter and comparison with the literature}
\label{sec:changeINb}

We now have all the ingredients necessary to evaluate the change in impact parameter, given in Eq.~\eqref{eq:EikonalImpactParameter} for general dressing, due to the soft dressing of initial states that connects the canonical and intrinsic frames. We will also explore how this deviation affects the mechanical angular momentum through Eq.~\eqref{eq:extraJloss}. 
Starting with the hard charge in scalar theory, QED, and gravity, given in equations \eqref{eq:LOHardChargeScalar}, \eqref{eq:LOHardChargeQED}, and \eqref{eq:LOHardChargeGR} respectively and inserting these expressions in Eq.~\eqref{eq:EikonalImpactParameter} yields 
\[
\left(x_{\perp}^\mu - b_{12}^\mu\right)^{\text{sc}} =& - \alpha(p_1+p_2)^\mu - \beta(p_1-p_2)^\mu \\
&
%- 
-\frac{e_1 e_2}{8\pi m_1 m_2} \left(-\frac{1}{6} \left(\frac{e_1 m^2_2}{ e_2 m^2_1} + \frac{e_2 m^2_1}{ e_1 m^2_2}  \right) -\frac{1}{\gamma^2 v^2} + \frac{ {\rm arctanh}(v)}{\gamma^2 v^3}\right) \Delta p^\mu+{\cal O}(\Delta p^3),
\]
for the scalar theory and
\[
\label{eq:QEDdeltaB}
\left(x_{\perp}^\mu - b_{12}^\mu\right)^{\text{qed}} =& - \alpha(p_1+p_2)^\mu - \beta(p_1-p_2)^\mu \\
&
%- 
+\frac{ e_1 e_2}{2\pi m_1 m_2}  \left(\frac{1}{3} \left(\frac{e_1 m_2}{ e_2 m_1} + \frac{e_2 m_1}{ e_1 m_2}  \right) -\frac{1}{\gamma v^2} + \frac{ {\rm arctanh}(v)}{\gamma^3 v^3}\right) \Delta p^\mu+{\cal O}(\Delta p^3),
\]
for QED, and
\[
\label{eq:GRdeltaB}
\left(x_{\perp}^\mu - b_{12}^\mu\right)^{\text{gr}}=&- \alpha(p_1+p_2)^\mu - \beta(p_1-p_2)^\mu\\
& 
%- 
-\frac{\kappa^2}{32\pi}\left(\frac{16}{3}-\frac{2}{v^2} + \frac{2(1-3 v^2){\rm arctanh}(v)}{v^3} \right)\Delta p^\mu+{\cal O}(\Delta p^3)
\]
for general relativity, to leading nontrivial order in the coupling. In these expressions we used the fact that $\sfq$ is the same as the impulse to leading nontrivial order. 

It is straightforward to extend these results to higher orders in the coupling. Proceeding to the next order, we include the hard charge contributions to the coefficients $\alpha$ and $\beta$ in Eq.~\eqref{eq:alphabeta} to obtain
\[
(x_\perp^\mu - b_{12}^\mu)_{Q,g^6} &= \frac{I(\gamma)}{m_1 m_2 (\gamma^2-1)}  \big(2 m_1 m_2 \left(\gamma ^2-1\right)\Delta p^\mu\\
&+u_1^\mu
   \left(\Delta p^2 (m_1 \gamma +m_2)+  2 m_1 m_2 (u_1\cdot
  \Delta p-\gamma   \, u_2\cdot\Delta p )\right)\\
  &+u_2^\mu
   \left(\Delta p^2 (m_1+m_2 \gamma )+2 m_1 m_2 (u_2\cdot\Delta p-\gamma 
    \, u_1\cdot\Delta p )\right)\big) \ ,
   \label{eq:NextOrderDeltaB}
\]
where $\gamma= p_1 \cdot p_2/m_1m_2$ as before, and
\[
I(\gamma)=
\begin{cases}
\dfrac{e_1 e_2}{16\pi m_1 m_2 }
\left[
-\dfrac{1}{6}\left(\dfrac{e_2 m_1^2}{e_1 m_2^2}
+\dfrac{e_1 m_2^2}{e_2 m_1^2}\right)
-\dfrac{1}{\gamma^2 v^2}
+\dfrac{\operatorname{arctanh}(v)}
{\gamma^2 v^3}
\right]
& \text{sc}, \\
-\dfrac{e_1 e_2}{4\pi m_1 m_2 }
\left[
\dfrac{1}{3}\left(\dfrac{e_2 m_1}{e_1 m_2}
+\dfrac{e_1 m_2}{e_2 m_1}
\right)-\dfrac{1}{\gamma v^2}
+\dfrac{\operatorname{arctanh}(v)}
{\gamma^3 v^3}
\right]
& \text{qed}, \\
\dfrac{\kappa^2}{64\pi}
\left[
\dfrac{16}{3}
-\dfrac{2}{v^2}
+\dfrac{2\left(1-3v^2\right)
\operatorname{arctanh}(v)}
{v^3}
\right]
& \text{gr} \ .
\end{cases}
\]

We may compare the leading order change in impact parameter due to the frame change in QED and in GR, given in Eqs.~\eqref{eq:QEDdeltaB} and \eqref{eq:GRdeltaB} respectively, with existing results obtained with traditional methods in Refs.~\cite{Damour:2020tta, Saketh:2021sri, Bini:2022wrq, DiVecchia:2022owy}. These references compute the leading radiated angular momentum in the intrinsic frame (i.e. the frame with a non-zero shear). Since this radiated angular momentum originates in the mechanical angular momentum of the matter system, it follows that the change of frame inducing the radiated angular momentum must also include a compensating change in the mechanical angular momentum, i.e. 
\[
J = p b \longrightarrow \delta J = p \delta b \ .
\]
It is indeed not difficult to see that the leading-order change in impact parameter in Eq.~\eqref{eq:GRdeltaB} reproduces the (negative of the) frame dependence in the angular momentum loss in general relativity in Eqs.~(5.22) and (6.24)-(6.26) of Ref.~\cite{Bini:2022wrq}, while the leading-order change in impact parameter in Eq.~\eqref{eq:QEDdeltaB} reproduces the frame dependence in the angular momentum loss in QED in Eqs.~(1.6)-(1.7) of Ref.~\cite{Saketh:2021sri}. 
The expression in Eq.~\eqref{eq:NextOrderDeltaB} provides a prediction for the frame-dependence of the angular-momentum loss at the next order.

\section{Conclusions}
\label{sec:conclusions}

In this work we developed an on-shell understanding of large gauge transformations, including supertranslations, by relating them to the non-uniqueness of massive charged asymptotic states in the presence of massless mediators.
A key ingredient was a family of coherent-state operators, built solely from zero-energy mediators, which can dress asymptotic massive charged particles and capture this non-uniqueness in the classical regime. 
We showed that each such coherent state defines a choice of asymptotic frame, i.e. a choice of time-independent expectation value of the mediator field, for scalar as well as for gauge mediators.  
The choice of coherent state is necessarily part of the intrinsic definition of the scattering states for charged particles\footnote{This parallels the familiar statement that charged operators require Wilson-line dressing to be well-defined under small gauge transformations~\cite{Giddings:2005id, Donnelly:2015hta, Cheung:2026euf}; the frame choice fixes the analogous freedom in the large-gauge sector.}.

In a strict on-shell framework, with in and out states defined at infinity, our coherent states are built from exactly-zero energy massless excitations and may be interpreted as resumming distribution-valued three-point $S$-matrix elements\footnote{This is somewhat in the spirit of Refs.~\cite{Guevara:2026qzd, Guevara:2026qwa} which discuss distribution-valued amplitudes localized on half-collinear configurations.}, as discussed in Ref.~\cite{Elkhidir:2024izo}.
In a more realistic scenario, with asymptotic states defined at large but finite distances $r$, the dressing is instead supported on frequencies $\omega \lesssim {\cal O}(1/r)$.

A central outcome of our analysis is that the argument of each coherent-state operator, which we refer to as the ``soft charge'', can be completed to a nonlocal operator that commutes with the scattering matrix, and is therefore conserved. 
We refer to the additional completion terms as the ``massive'' and ``massless'' hard charges, according to whether they involve only massive or massless external particles, respectively. 
The conservation of the complete charge operator, i.e. of the sum of the soft and hard charges, implies that amplitudes with one or more external massless particles of exactly zero energy are related, by a \emph{finite} rescaling determined by the hard charge, to amplitudes in which the zero-energy states are absent.
Our construction shows that the amplitudes with zero energy states are distributionally supported on a distinguished subspace of momentum space and are not merely limits of generic-kinematics configurations. It moreover shows that the contribution of this locus is not independent: as mentioned, it is related to a finite rescaling of the amplitude away from that locus, thereby mapping the evolution of the precisely zero-energy sector to the evolution of the finite-energy one.
In other words, the hard content of the final state is obtained by evolution within the hard Hilbert space alone, with the initial zero-energy data whose effect within the hard sector reduces to the kinematic rephasing by the hard-charge eigenvalue.
This phase cancels in inclusive observables and  
is absorbed by reparametrization of local observables so the apparent non-decoupling of the hard and zero-energy modes is in fact innocuous. 

These results sharpen the physical meaning of dressing asymptotic states with zero-energy quanta: it organizes the freedom in defining charged asymptotic states, implements the action of asymptotic symmetry generators, and resums distributionally-supported scattering amplitudes with zero-energy external mediators. The exactly-zero-energy sector enters as intrinsic asymptotic data rather than as a boundary of generic kinematics.~\footnote{This is distinct from the traditional use of coherent-state dressings, beginning with Kulish and Faddeev, which aims at taming the infrared divergences of scattering amplitudes, see e.g.~\cite{Carney:2018ygh, Lippstreu:2025jit, Chicherin:2025keq}. 
Our dressing in  gauge and gravity theories involves a projector that removes low spherical harmonics ($\ell=0$ in gauge theories and $\ell=0,1$ in gravity) of the transformation parameter, which are the ones contributing to IR divergences. The two types of dressing are complementary and can in principle be combined.}

%%%%%%%%%%%%%%%%%%%%

In theories where asymptotic symmetries admit a geometric interpretation, such as gravitational theories in asymptotically-Minkowski spacetimes, these conserved operators are naturally identified with the generators of the corresponding asymptotic symmetries. From this viewpoint, amplitudes with exactly zero-energy external particles encode the action of these symmetries on the scattering operator, while the dressing ambiguity determines how that action is realized on the asymptotic states. 
The dynamics of the zero-energy sector is captured by amplitudes supported on special kinematic configurations, whose physical content is given by the action of the soft charge on the scattering operator.

Our construction holds in all theories with charged massive particles and massless mediators, demonstrating that the existence of local symmetries in the theory is not a prerequisite for the existence of an asymptotic symmetry. 
An illustration is provided by theories with scalar mediators and trilinear couplings, see Sec.~\ref{sec:scalars}. While these theories have no local symmetries,~\footnote{See Refs.~\cite{Campiglia:2018see,Francia:2018jtb} for discussions of theories with massless scalars that have local (shift) symmetries; our construction does not rely on such a structure.} 
they nevertheless exhibit the same structure that in gauge and gravity theories descends from one: an ambiguity in the definition of massive charged states whose ``large'' part acts nontrivially on asymptotic data and is generated by conserved charges. This is manifest in our construction, where $Q_h$ is fixed by the requirement that $[S, Q_s + Q_h] = 0$, without reference to a pre-existing local symmetry, and our scalar-mediator results provide a concrete realization of the corresponding asymptotic symmetry.
This moreover supports the idea that amplitude soft factorization in the soft limit implies the existence of a conservation law.

We also showed that the dressing generates  a finite energy-dependent shift in the impact parameter. 
Starting from the exact calculation of the hard charge $Q_h$ for an arbitrary gauge parameter $T(n)$, we obtained a closed-form expression for this shift, valid to all orders in perturbation theory. In gravity, this makes transparent how the dressing encodes the BMS-frame dependence of angular momentum and ensures that the total angular momentum is frame-independent.
For the intrinsic $\leftrightarrow$ canonical BMS frame relation, we reproduced the known leading-order change in impact parameter due to radiation reaction~\cite{Bini:2022wrq}, interpreting it as a kinematic effect of the dressing. 
We also reproduced the analogous results in QED~\cite{Saketh:2021sri}.
Expanding our general all-order expression yields a novel prediction for the angular momentum loss and for the shift in the impact parameter
that compensates it at the next-to-leading order in Newton's constant.

The scalar field example suggests taking the asymptotic action $Q$ as fundamental (and not its symmetry origin, which may or may not be apparent). 
In QED, the exponentiated soft operator $e^{iQ_s[T]}$ can be viewed as the zero-frequency part of a Wilson-line dressing, with the relative transformation between two distinct such dressings defining a Wilson loop~\cite{Choi:2018oel}. 
This connects the construction to generalized symmetries. For example, in pure Maxwell theory Wilson loops are charged line operators of the electric one-form symmetry. Dynamical electrically charged matter breaks this symmetry in the bulk by rendering its codimension-two flux operators non-topological, although a related angle-dependent structure may survive as an asymptotic symmetry~\cite{Gaiotto:2014kfa,Lake:2018dqm}.
\footnote{See also \cite{Tizzano:2026rgr} for a discussion of connections between symmetry operators, asymptotic charges and soft theorems in QED.}
The impact-parameter shift found above gives a concrete hard-sector realization of the same asymptotic action on matter data: the momentum-dependent phase generated by $Q_h[T]$ translates the transverse coordinate conjugate to the momentum transfer.
Finally, since $T(n)$ acts ultralocally on celestial data (e.g. in gravity, as independent translations along each null generator) the family $Q[T]$ is  kinematically reminiscent of subsystem symmetries~\cite{Seiberg:2020bhn}, though the interacting dynamics freely redistributes charge among generators at zero soft-energy cost. Identifying the corresponding boundary operators and their algebra or fusion would determine the precise generalized-symmetry structure.

Throughout our analysis we assumed that the soft dressing is independent of the retarded time. 
This assumption was embodied in Eq.~\eqref{localization} and reflects the fact that the dressing captures the asymptotic field of initial-state particles located strictly in the infinite past. 
As discussed in Sec.~\ref{sec:basicSetup}, in more realistic scenarios, the initial-state data is located at some large but finite past time. The corresponding initial shear need not vanish, and will in general be 
$u$-dependent: locating the initial data at finite past time $T$ effectively allows the dressing phase to be supported on $\omega \lesssim {\cal O}(1/T)$.
Furthermore, initial-state data located at ${\cal I}^-$ include incoming finite-energy radiation, described by a background field $C_{AB}(u, n)$.
The supertranslations of such a background include a term proportional to the time derivative of the shear, cf. Eq.~\eqref{supertranslation}. 
We can accommodate such initial conditions simply by modifying the shape functions which define the coherent state to include the term (recall that $k = \omega n$)
\[
 \delta F^\helicity_p(k)  = 
T_p(n) \,
{\bar\epsilon}{}^\helicity{}^{AB}
 \, \int \dd u \, \partial_u C_{AB} \, e^{+i\omega u} \ .
\] 
Whether the full construction, in particular completion of the argument of the dressing phase to a conserved charge, survives beyond the linearized approximation is an interesting question which we leave for future work.

A noteworthy by-product of our analysis is the close relationship, established in Sec.~\ref{sec:VVKS}, between scattering amplitudes and spacetimes in Kerr-Schild form. Scattering-amplitude computations, with the standard choice of in-state for which $\langle \psi|a_\eta(k)|\psi\rangle = 0$ including for the zero-energy modes, are naturally set up in the canonical BMS frame. On the classical side, the canonical frame is the one whose asymptotic data coincides with that of each scattering body's metric in Kerr-Schild coordinates. This identification, which holds for a single boosted Schwarzschild\footnote{It would undoubtedly be interesting to extend the analysis in this paper to spinning bodies.} black hole and, with the qualifications discussed in Sec.~\ref{sec:VVKS}, for a collection of black holes, may shed further light on why Kerr-Schild coordinates are so natural for the classical double copy~\cite{Monteiro:2014cda}. 
It also raises the possibility that organizing classical perturbation theory directly around Kerr-Schild data, rather than passing through the De~Donder gauge and a subsequent supertranslation to the canonical frame, could yield a simpler classical formalism for scattering observables. We leave this question for future work.

We finish with some thoughts on possible extensions of our analysis.
%It would be interesting to extend our analysis in several directions. 
%
One such direction concerns the extent to which the present construction generalizes beyond coherent-state dressings supported on the exactly zero-energy locus (cf. the finite-distance and time-dependent extensions above) and beyond the cases where asymptotic symmetries are geometrically manifest. 
It was demonstrated in Ref.~\cite{DeAngelis:2025vlf} that departing from the strict asymptotic limit in the presence of certain infrared singularities introduces a breakdown of the peeling property and may interfere with the existence of Bondi coordinates. While further investigation is necessary to determine whether BMS supertranslations survive as a symmetry in such settings, the field-theoretic construction described here appears to be insensitive to these issues and may yield nonlocal conservation laws even when no geometric counterpart exists.
Another direction is to understand more fully the role of self-interacting mediators and the consequences of a comparable coherent-state dressing of finite-energy mediators, $\omega\gg 1/r, 1/T$, describing finite-energy radiation in the initial state.
More broadly, our results suggest that the long-range field of a particle, its asymptotic state, and the distributional support of the scattering matrix at zero energy are three presentations of the same data. In this picture, infrared structure is part of the definition of the asymptotic states; we expect that this perspective will be useful well beyond the gravitational setting.

\section*{Acknowledgements}

We thank Tim Adamo, Zvi Bern, Thomas Dumitrescu, Carlo Heissenberg, Enrico Herrmann, Gary Horowitz, Anton Ilderton, David Kosower, James Lucietti, Ricardo Monteiro and Michael Saavedra.
DOC is supported by the STFC grant ``Particle Theory at the Higgs Centre'' and by the European Research Council under
Advanced Investigator grant ERC–AdG–101200505.
RR~is supported by the U.S.  Department of Energy (DOE) under award number~DE-SC00019066 and by a Senior Fellowship at the Institute for Theoretical Studies, ETH Z\"urich.
For the purpose of open access, the author has applied a Creative Commons Attribution (CC BY) licence to any Author Accepted Manuscript version arising from this submission.

\newpage

\appendix

\section{Rudiments of BMS 
}\label{app:BMS}

In this appendix, we review some aspects of BMS transformations that will be useful for our definition of the corresponding dressing of the single-particle Fock states.  To this end we will follow Ref.~\cite{Flanagan:2015pxa}, up to the change in signature from mostly-plus to mostly-minus, which has been our convention for the main text of the paper.

The BMS symmetry is defined as the subset of large diffeomorphisms which preserve the Bondi gauge condition
\begin{align}
\label{eq:Bondigauge}
g_{rr}=0
~,~~~
g_{rA}=0
~,~~~
\partial_r \det\gamma_{AB} = 0 \ ,
\end{align}
which puts the metric in the form 
\begin{align}
\dd s^2 = U e^{2\beta} \dd u^2 + 2 e^{2\beta} \dd u \, \dd r
- r^2 \gamma_{AB} 
(\dd\theta^A - {\cal U}^A \dd u)
(\dd\theta^B - {\cal U}^B \dd u) \ ,
\end{align}
with the large distance fall-off
\begin{align}
\label{eq:coefexpansion}
\gamma_{AB}&=\Omega_{AB} +\frac{1}{r} C_{AB} + \frac{1}{r^2} \left(\frac{\Omega_{AB}}{4} C^{CD}C_{CD} +{\cal D}_{AB} \right)+{\cal O}(r^{-3})
\\
{\cal U}^A &= \frac{1}{r^2} U^A+{\cal O}(r^{-3})
~,\quad
\beta = \frac{\beta_0}{r} + \frac{\beta_1}{r^2}+{\cal O}(r^{-3})
~,\quad
U = 1-\frac{2m}{r} -  \frac{2\cal M}{r^2} +{\cal O}(r^{-3}) \,.
\nonumber
\end{align}
Imposing the Bondi gauge implies that the shear $C_{AB}$ is traceless with respect to the round $S^2$ metric $\Omega_{AB}$, and that ${\cal D}_{AB}$ is traceless, and structurally similar conditions on the coefficients of $r^{-(n\ge3)}$ in the expansion of $\gamma_{AB}$. \footnote{We use here a slightly different normalization of the shear tensor than the one used in Ref.~\cite{Elkhidir:2024izo}. }

Einstein's equations imply that
\[
U_A = -\frac{1}{2} D^B C_{AB}
~,\quad
\beta_0=0
~,\quad
\beta_1=-\frac{1}{32} C^{AB}C_{AB} -
\text{matter contrib.}
\]
\[
\dot m = -\frac{1}{8} \partial_u C_{AB}\, \partial_u C^{AB} +\frac{1}{4} D_A D_B \partial_u C^{AB} +\text{(matter contrib.)}
\]
where the last relation is the celebrated Bondi mass loss formula, usually written in terms of the news tensor $N^{AB}=\partial_u C^{AB}$. The index contraction is taken with the 
Euclidean round $S^2$ metric.

The BMS transformations in the sense defined above, as the transformations preserving the Bondi gauge condition~\eqref{eq:Bondigauge} and the falloff of coefficients in Eq.~\eqref{eq:coefexpansion}, are generated by the vector field
\[
\vec \xi = \xi^u \partial_u + \xi^A\partial_A =
         \left(T(n)+\frac{1}{2} u D_A Y^A(n)\right)\partial_u +Y^A(n)\partial_A \ ,
\]
where, as in the main text, $n=(1, \bm n)$ and $\bm n$ is the unit vector parametrizing $S^2$.
The supertranslation, defined as the transformation shifting the retarded time, is accompanied by a suitable change in the coefficients of the large-$r$ expansion of the metric:
\[
\delta u &= T(n)
~,\quad
\delta C_{AB} = T(n) \, \partial_u C_{AB} + \left(\Omega_{AB} D^2T(n) -2 D_A D_B T(n) \right) \,,
\\
\delta m &= T(n)\, \dot m +\frac{1}{4} \partial_u C^{AB} D_AD_BT(n) 
+\frac{1}{2} D_AT(n) \, D_B \partial_u C^{AB} \ .
\label{supertranslation}
\]
These transformations are also generated by the N\"other charge
\[
Q_T = \int  \dd^2 \Omega\,  m \, T(\theta,\phi) \ ,
\]
where $m$ is the Bondi mass aspect, see Eq.~\eqref{eq:coefexpansion}.

That is, BMS supertranslations change the shear, i.e. the transverse-traceless part of the angular ${\cal O}(1/r)$ part of the metric (as well as other components of the metric at the same order in $1/r$, as dictated by Einstein's equations, except that we cannot access them with on-shell methods). As we discussed in Sec.~\ref{sec:frame}, a soft dressing of single Fock-particle states also changes the asymptotic metric.

%======================================================================
\section{Integrals for general dressing}
\label{app:int_general_dressing}
%======================================================================

In this Appendix we evaluate the $l$ integral $I_0$ and $I_P$ needed in Sec.~\ref{sec:hard_massless_general} and Sec.~\ref{sec:hard_matter_general}, respectively,
\[
I^0_{p,K} 
&= \int \dd \Phi(l) \hdelta(\omega_l) \frac{ K^A K^B \, (\Omega_{AB}D^2-2D_A D_B) T_p(n_l)}{2 K \cdot l + i \epsilon} \ ,
\\
I_{p,P} 
&= \int \dd \Phi(l) \hdelta(\omega_l) \frac{ P^A P^B \, (\Omega_{AB}D^2-2D_A D_B) T_p(n_l)}{2 P \cdot l + i \epsilon} \ ,
\label{app:integrals_defs}
\]
where the on-shell condition in the phase space measure sets $l^\mu = \omega_l n_l^\mu(\theta, \phi)\equiv \omega_l n_l$ with $n_l^2 = 0$. The vectors $K$ and $P$ obey on-shell conditions $K^2=0$, $P^2 = m^2$,  $K^A = K^\mu e(l)_\mu^A = K^\mu D^A (n_l)_\mu$ with $D^A$ the covariant derivative on the 2-sphere, and similarly for $P^A$. Last but not least, in this appendix we will drop the index $p$ on $T_p$.

We will compute the latter integral, and obtain the former as the massless limit. To this end we collectively denote $K$ and $P$ as $K$, and normalize it as
\[
K = (1, \Kvec) \ ,
\]
with $\Kvec^2 = 1$ for the former and $\Kvec^2 <1$ for the latter; for the purpose of the integral, $\Kvec$ is a fixed vector in $\mathbbm{R}^3$. We also write 
\[
n_l = (1, \nhat) \ ,
\]
and the measure sets $\nhat^2 = 1$, i.e. $\nhat$ parametrizes $\Sphere$, with round sphere metric $\Omega^{AB}$ and Levi-Civita connection $D_A$. 

For later convenience we define the norm of $\Kvec$ and
the unit vector along it
\begin{equation}\label{eq:kappadef}
\nK \equiv |\Kvec|,\qquad \Khat \equiv \frac{\Kvec}{\nK}
~~,\quad(\nK>0) \ .
\end{equation}
At each point $\nhat\in\Sphere$ we define the tangential projection
\begin{equation}\label{eq:Ktan}
K^i_\parallel = K^i - (\Kvec\cdot\nhat)\,n^i  \ ,
\end{equation}
identified with the intrinsic vector $K^A$ on $\Sphere$. We further define
\begin{equation}\label{eq:ucdef}
\xi \equiv \Kvec\cdot\nhat = \nK\,c,\qquad c\equiv\Khat\cdot\nhat \ .
\end{equation}

With these preparations, and noting that with the parametrizations above the complete $\omega_l$ dependence factors out so the integral over $\omega_l$ is straightforwardly evaluated using the delta function, we are 
left to evaluate 
\begin{equation}\label{eq:theintegral}
I_{p,K} = \frac{1}{2}\times\frac{1}{2}\times \frac{1}{(2\pi)^2}\int_{\Sphere} d^2\Omega_{\bm n}\;
\frac{K^A K^B (\Omega_{AB}D^2-2D_A D_B)\,T_p(n)}{1-\xi } \ ,
\end{equation}
where we slightly adjusted the notation here for the argument of $T$. We also pedantically exposed the origin of the numerical factor: the $(2\pi)^2$ is the remnant of the $(2\pi)$ factors in the definition of the measure and of $\hdelta$, the first $1/2$ comes from the $1/(2\omega)$ in $\dd\Phi(l)$ and the second factor of $1/2$ is explicitly in the denominator of Eq.~\eqref{app:integrals_defs}.
The plan is to integrate by parts; we will therefore begin by deriving various identities involving derivatives of $K^A$ and $u$.

\subsection{Some identities}\label{sec:identities}

\noindent
1) Norm of $K^A$ on $\Sphere$:
\begin{equation}
\Omega_{AB}\,K^A K^B = \nK^2 - \xi^2 \ .
\end{equation}
The proof amounts to using the definition of $K^A$ as the projection of $K$ on $\Sphere$:
\begin{align}
\Omega_{AB}\,K^A K^B &= \delta_{ij}\,K^i_\parallel\,K^j_\parallel
= (K^i - \xi\,n^i)(K^i - \xi\,n^i)\nonumber\\
&= |\Kvec|^2 - 2\xi(\Kvec\cdot\nhat) + \xi^2|\nhat|^2
= \nK^2 - 2\xi^2 + \xi^2 = \nK^2 - \xi^2 \ .
\label{eq:normsq}
\end{align}

\noindent
2) The derivative of $\xi = \Kvec\cdot \nhat$:
\begin{equation}
D_A \xi = K_A \ .
\end{equation}
Since $\xi = \delta_{ij} K^i n^j$ and $\Kvec$ is constant, the ambient gradient is $\partial_i \xi = K^i$. 
The intrinsic gradient is its tangential projection:
$D_A \xi = e_A^i\,K^i = K_A$,
where $e_A^i = \partial x^i/\partial z^A$ are the tangent vectors of the embedding.

\noindent
3) The $\Sphere$ covariant derivative of the projection of $\Kvec$: 
\begin{equation}
D_A K_B = -\xi\,\Omega_{AB} \ .
\end{equation}
To prove this we use that $K_A = e_A^i K^i$ with $e_A^i = \partial x^i/\partial z^A$, so
$D_A K_B = (D_A e_B^i)\,K^i$.
The Gauss--Weingarten equation for the unit sphere states
\begin{equation}\label{eq:GW}
D_A e_B^i = -\mathcal{K}_{AB}\,n^i \ ,
\end{equation}
where $\mathcal{K}_{AB}$ is the second fundamental form. For the unit sphere 
$\mathcal{K}_{AB}=\Omega_{AB}$, giving
\begin{equation}
D_A K_B = -\Omega_{AB}\,n^i K^i = -\xi \,\Omega_{AB} \ .
\label{eq:DaKb}
\end{equation}
From here, by contracting with $\Omega^{AB}$ and $K^A$, respectively, it also follows that 
\begin{align}
D_A K^A &= -2\xi \ , \label{eq:divK} 
\\
K^A D_A K^B &= -\xi \,K^B \ .
\label{eq:KDK} 
\end{align}

\noindent
4) The derivatives of relevant functions of $\xi$:
\begin{align}
K^A D_A h(\xi ) & = K^AK_A  \frac{dh(\xi )}{d\xi }  = (\nK^2-\xi^2)\frac{dh(\xi )}{d\xi } \ ,
\\
K^A D_A \frac{1}{1-\xi } &= \frac{\nK^2-\xi^2}{(1-\xi )^2} \ .
\end{align}

To proceed with the evaluation of \eqref{eq:theintegral} by integrating by parts it is convenient to first slightly reorganize the integrand as
\begin{align}
K^A K^B D_A D_B T_p(n) = (K^A D_A) (K^B D_B)T_p(n) + \xi \,(K^A D_A) T_p(n) \ ,
\end{align}
which can be proven by applying the Leibniz rule and \eqref{eq:DaKb}. Then,
\begin{align}
\frac{K^A K^B D_A D_B T}{1-\xi } &= \frac{(K^A D_A) (K^B D_B)T +  (K^A D_A) T}{1-\xi } - (K^A D_A) T \equiv I_1+I_2
\\
\frac{-\frac{1}{2}K^A K^B \Omega_{AB}\, D^2 T}{1-\xi } &= \frac{-\frac{1}{2}(\nK^2-1)\, D^2T}{1-\xi } -\frac{1}{2}(1+\xi )\, D^2T   \equiv I_3+I_4 \ ,
\end{align}
so the integral to be evaluated is a sum of four parts:
\begin{align}
I_{p,K} = \frac{-1}{8\pi^2}(I_1+I_2+I_3+I_4) \ .
\end{align}
If $\nK<1$, then $\xi <1$ so $\xi -1<0$; however, we are interested also in taking the limit $\nK\to 1$, so we will need 
to keep track of the terms that become singular when $\xi \to 1$. 

\subsection{Regularized integration by parts}

To keep track of the terms that become singular when $\xi\to 1$ we will excise a small neighborhood of $\Sphere$ around $u=1$, $\Sphere\to \Sphere\backslash C_\epsilon$; for $\nK<1$ this excision is not part of the integration domain and the limit $\epsilon\to 0$ is trivial.
For $\nK=1$ however it will be important. The order in which the limit is taken is \emph{first} $\nK\rightarrow 1$ and \emph{afterwards} $\epsilon\rightarrow 0$.

For arbitrary $f$ and $g$ functions, the integration by parts rule is:
\begin{align}
\int_{\Sphere\backslash C_\epsilon} d^2\Omega_{\bm n} \, g K^A D_A f = - \int_{\Sphere\backslash C_\epsilon} d^2\Omega_{\bm n} 
\left[g f D^A K_A + f K^A D_A g \right] +  \int_{\partial C_\epsilon} ds\, K^A m_A \, g \, f \ ,
\end{align}
where $ds$ is the line element on $\partial C_\epsilon$ and $m_A$ is the outward-pointing unit vector inside $\Sphere$ and normal to the boundary  $\partial C_\epsilon$.

\subsection{Evaluation of integrals $I_1,\dots I_4$}

The integrals $I_2$ and $I_4$ have no potential singularities. We may therefore discard the regularization from the outset; they evaluate to
\begin{align}
I_2 &:= -\int d^2\Omega_{\bm n} K^A D_A T_p(n) =  \int  d^2\Omega_{\bm n} D_A K^A T_p(n) =-2 \int d^2\Omega_{\bm n} \, \xi \, T_p(n) 
\\
I_4 &:= -\frac{1}{2} \int d^2\Omega_{\bm n} \, (1+\xi )D^2 T_p(n) = -\frac{1}{2} \int d^2\Omega_{\bm n}\, D^2 \xi \, T_p(n) = \int d^2\Omega_{\bm n} \, \xi \, T_p(n) =-\frac{1}{2} I_2
\end{align}

For integrals $I_1$ and $I_3$ we integrate by parts once and get
%%%%%%%%%%%%%%%%%%%%%%%%%%%
\begin{align}
I_1 :=& \int_{\Sphere\backslash C_\epsilon} d^2\Omega_{\bm n} \frac{1}{1-\xi }(K^A D_A)\left[ (K^B D_B)T_p(n) +   T_p(n) \right]
\cr
=& I_2 -\int_{\Sphere\backslash C_\epsilon} d^2\Omega_{\bm n} \,   T_p(n)   -(\nK^2-1)I_\text{aux}
 -(\nK^2-1)\int_{\Sphere\backslash C_\epsilon} d^2\Omega_{\bm n} \frac{T_p(n)}{(1-\xi )^2}  
\cr
&     +  \int_{\partial C_\epsilon} ds\,
\frac{K^A m_A}{1-\xi }\left[ (K^B D_B)T_p(n) +   T_p(n) \right] \ ,
\\
I_3 :=& -\frac{1}{2}(\nK^2-1) \int_{\Sphere\backslash C_\epsilon} d^2\Omega_{\bm n} \frac{D^2 T_p(n)}{1-\xi } 
\cr
= &
+\frac{1}{2}(\nK^2-1) I_\text{aux} 
-\frac{1}{2}(\nK^2-1) \int_{\partial C_\epsilon} ds \, \frac{m^A D_A T_p(n)}{1-\xi } \ ,
\end{align}
The auxiliary integral is
\begin{align}
I_\text{aux} &:= \int_{\Sphere\backslash C_\epsilon} d^2\Omega_{\bm n} \frac{(K^B D_B)T_p(n) }{(1-\xi )^2}  =
- \int_{\Sphere\backslash C_\epsilon} d^2\Omega_{\bm n} \frac{2\nK^2  - 2 \xi }{(1-\xi )^3} T_p(n)
+\int_{\partial C_\epsilon} ds\, \frac{m_A K^A \, T_p(n) }{(1-\xi )^2} 
\cr
&= 
- 2(\nK^2-1)\int_{\Sphere\backslash C_\epsilon} d^2\Omega_{\bm n} \frac{T_p(n) }{(1-\xi )^3}  - 2\int_{\Sphere\backslash C_\epsilon} d^2\Omega_{\bm n} \frac{T_p(n) }{(1-\xi )^2}
+\int_{\partial C_\epsilon} ds\, \frac{m_A K^A \, T_p(n) }{(1-\xi )^2}  \ ,
\end{align}
where in the second equal sign we integrated by parts one more time.

Putting together $I_{1,2,3,4}$, we obtain
\begin{eqnarray}
-8\pi^2 \, I_{p, K}  
%         \\
&=& -\int_{\Sphere\backslash C_\epsilon} d^2\Omega_{\bm n} \,  \left(1+3 \xi \right) T_p(n)     +  \int_{\partial C_\epsilon} ds\,\frac{K^A m_A}{1-\xi }\left[ (K^B D_B)T_p(n) +   T_p(n) \right]
\cr
&&+(\nK^2-1)^2 \int_{\Sphere\backslash C_\epsilon} d^2\Omega_{\bm n} \frac{T_p(n) }{(1-\xi )^3} 
\\
&&         -\frac{1}{2}(\nK^2-1) \int_{\partial C_\epsilon} ds \, \frac{m^A D_A T_p(n)}{1-\xi }
          -\frac{1}{2}(\nK^2-1) \int_{\partial C_\epsilon} ds\, \frac{m_A K^A \, T_p(n) }{(1-\xi )^2} \ .
\nonumber
\end{eqnarray}

For $\nK\ne 1$ all boundary integrals can be dropped.
In the massless limit, in which the sequence of limits starts with $\nK\to 1$,  only the first line survives, so it is only the first boundary integral that we need to evaluate.
Consequently, the last line can always be dropped and, effectively, the integral for general $\nK$ is
\begin{eqnarray}
-8\pi^2 I_{p, K}  
&=& -\int_{\Sphere\backslash C_\epsilon} d^2\Omega_{\bm n} \,  \left(1+3 \xi \right) T_p(n)    +  \int_{\partial C_\epsilon} ds\,\frac{K^A m_A}{1-\xi }\left[ (K^B D_B)T_p(n) +   T_p(n) \right]
\cr
&&+(\nK^2-1)^2 \int_{\Sphere\backslash C_\epsilon} d^2\Omega_{\bm n} \frac{T(\nhat) }{(1-\xi )^3} 
\end{eqnarray}
We note here that the first term picks out of $T_p(n)$ the terms proportional to the $Y_{00}$ and $Y_{1m}$ spherical harmonics. However, those terms are projected out of $T_p(n) $ by the operator $(\Omega_{AB}D^2-2D_A D_B)$ in the original integrand. So these terms can also be ignored.

\subsection{Evaluation of the boundary integral(s)}

Since the regularization excises a small neighbourhood of the point $\Khat$ on $\Sphere$, we parametrize $\nhat$  relative to $\Kvec$; that is, we can effectively assume that $\Kvec$ points in the $\hat z$ direction and use polar coordinates for $\nhat$. 
The boundary of $C_\epsilon$ is then parametrized by $\phi\in[0, 2\pi)$ placed at $\theta=\epsilon$. In these coordinates $\xi$ is
\begin{equation}
\xi = \Khat\cdot \nhat = \cos\epsilon \ ,
\end{equation}
i.e. in these coordinates $\nhat|_{\theta=0} = \Khat$. 
The line element is 
\begin{equation}
ds = \sin\epsilon \, d\phi \ ,
\end{equation}
and the unit vector $m$ points in the negative $\theta$ direction
\begin{equation}
m_A dx^A = -d\theta 
\qquad
K^Am_A = (-\sin\theta)(-1) \;\bigg|_{\theta=\epsilon} = \sin\theta\;\bigg|_{\theta=\epsilon} = \sin\epsilon
\ .
\end{equation}
Since $T_p(n)$ is a scalar function, we may replace the covariant derivative acting on it by a partial derivative 
\begin{equation}
(K^B D_B)T_p(n) = (K^B \partial_B)T_p(n) = -\sin\theta \, \partial_\theta T_p(n) \ ,
\end{equation}
where in the second equality we used that the coordinates are relative to $\Khat$.

Thus, the boundary integral is
\begin{align}
&\lim_{\epsilon\to 0} \int_{\partial C_\epsilon} ds\,\frac{K^A m_A}{1-\xi }
\left[ (K^B D_B)T_p(n) +   T_p(n) \right] 
\\
&= 
\lim_{\epsilon\to 0}
\int_0^{2\pi} d\phi\, \sin\epsilon \frac{\sin\epsilon}{1-\cos\epsilon} \left[  -\sin\theta \, \partial_\theta T_p(n)  +   T_p(n) \right] \;\bigg|_{\theta=\epsilon} = 4\pi T_p(\Khat)= 4\pi T_p(\Kvec) \ ,
\nonumber
\end{align}
where in the last equal sign we used the fact that this integral is evaluated in the limit $\Khat$ has unit norm, $\nK=1$ so $\Kvec = \Khat$.

Thus, the integral is:
\begin{align}
I_{p, K}  =& -\delta_{1, \nK} \, \frac{T_p(\Kvec)}{2\pi} 
%-\int d^2\Omega_{\bm n} \,  \left(1+\frac{3}{2} u\right) T(\nhat)     
-\frac{1}{2\pi}(\nK^2-1)^2 \int_{\Sphere } 
\frac{d^2\Omega_{\bm n}}{4\pi} \frac{T_p(n) }{(1-\xi )^3} \ ,
\label{eq:integral_final}
\end{align}
where we also dropped the terms proportional to the $Y_{00}$ and $Y_{1m}$ components of $T_p(n)$.
We use this expression in Secs.~\ref{sec:hard_massless_general} and \ref{sec:hard_matter_general}.

\newpage

%\bibliographystyle{abbrvurl}
%\bibliographystyle{JHEP}
%\bibliography{biblio}{}

\providecommand{\href}[2]{#2}\begingroup\raggedright\endgroup

\end{document}